\documentclass[longauth]{aa} 

\graphicspath{{./img/}}
\usepackage{graphicx}
\usepackage{caption}
\usepackage{subcaption}
\usepackage{float}
\usepackage{comment}
\usepackage{natbib}
\usepackage{txfonts}
\usepackage{xcolor}
\usepackage[symbol]{footmisc}

\newcommand{\cii}{\mbox{C\,{\sc ii}}} 
 
\newcommand{\sii}{\mbox{S\,{\sc ii}}} 
 
\newcommand{\siii}{\mbox{Si\,{\sc ii}}} 
 
\newcommand{\siiv}{\mbox{Si\,{\sc iv}}} 
\newcommand{\oi}{\mbox{O\,{\sc i}}} 

\newcommand{\siiis}{\mbox{Si\,{\sc ii$^{\star}$}}} 
 
\newcommand{\ciis}{\mbox{C\,{\sc ii$^{\star}$}}}

\usepackage{hyperref}
\hypersetup{%
    colorlinks=true, 
    citecolor=blue,
    filecolor=blue,
    linkcolor=blue,
    urlcolor=black
    }

\begin{document} 

   \title{The most distant optically polarised GRB afterglow:\\GRB 240419A at {\slshape z} = 5.178}

   \author{R.~Brivio\inst{1}\fnmsep\thanks{email: riccardo.brivio@inaf.it}
           \and
           S.~Covino\inst{1,2}
           \and
           M.~Ferro\inst{1}
           \and
           A.~Saccardi\inst{3}
           \and
           A.~Martin-Carrillo\inst{4}
           \and
           A.~Kuwata\inst{5,6}
           \and
           K.~Toma\inst{5,6}
           \and
           P.~D'Avanzo\inst{1}
           \and
           Y.-D.~Hu\inst{7,1}
           \and
           L.~Izzo\inst{8,9}
           \and
           S.~Kobayashi\inst{10}
           \and
           T.~Laskar\inst{11}
           \and
           G.~Leloudas\inst{12}
           \and
           D.~B.~Malesani\inst{9,13,14}
           \and
           M.~Pursiainen\inst{15}
           \and
           S.~Vergani\inst{16}
           \and
           K.~Wiersema\inst{17}
           \and
           S.~Bloemen\inst{14}
           \and
           S.~Campana\inst{1}
           \and
           V.~D'Elia\inst{18}
           \and
           S.~de Wet\inst{12}
           \and
           M.~de~Pasquale\inst{19}
           \and
           P.~J.~Groot\inst{14,20,21,22}
           \and
           P.~Jakobsson\inst{23}
           \and
           J.~Mao\inst{24}
           \and
           A.~Melandri\inst{25}
           \and
           G.~Pugliese\inst{26}
           \and
           A.~Rossi\inst{27}
           \and
           R.~Salvaterra\inst{28}
           \and
           B.~Schneider\inst{29,30}
           \and
           N.~R.~Tanvir\inst{31}
           \and
           J.~van~Roestel\inst{26}
           \and
           P.~M.~Vreeswijk\inst{14}
           \and
           T.~Zafar\inst{32}
          }

   \institute{INAF – Osservatorio Astronomico di Brera, Via E. Bianchi 46, 23807 Merate (LC), Italy
              \and
              Como Lake centre for AstroPhysics (CLAP), DiSAT, Università dell’Insubria, via Valleggio 11, 22100 Como, Italy
              \and
              Université Paris-Saclay, Université Paris Cité, CEA, CNRS, AIM, 91191, Gif-sur-Yvette, France
              \and
              School of Physics and Centre for Space Research, University College Dublin, Belfield, D04 V1W8 Dublin, Ireland
              \and
              Astronomical Institute, Graduate School of Science, Tohoku University, Sendai 980-8578, Japan
              \and
              Frontier Research Institute for Interdisciplinary Sciences, Tohoku University, Sendai 980-8578, Japan
              \and
              Guangxi Key Laboratory for Relativistic Astrophysics, School of Physical Science and Technology, Guangxi University, Nanning 530004, China
              \and
              INAF - Osservatorio Astronomico di Capodimonte, Salita Moiariello 16, 80131 Napoli, Italy
              \and
              Cosmic Dawn Center (DAWN), Copenhagen, Denmark
              \and
              Astrophysics Research Institute, Liverpool John Moores University, Liverpool Science Park IC2, 146 Brownlow Hill, Liverpool, L3 5RF, UK
              \and 
              Department of Physics \& Astronomy, University of Utah, Salt Lake City, UT 84112, USA
              \and
              DTU Space, Technical University of Denmark, Elektrovej 327-328, DK-2800 Lyngby, Denmark
              \and
              Niels Bohr Institute, University of Copenhagen, Jagtvej 128, 2200 Copenhagen N, Denmark
              \and
              Department of Astrophysics/IMAPP, Radboud University, 6525 AJ Nijmegen, The Netherlands
              \and
              Department of Physics, University of Warwick, Gibbet Hill Road, Coventry CV4 7AL, UK
              \and
              LUX, Observatoire de Paris, Université PSL, Sorbonne Université, CNRS, 92190 Meudon, France
              \and
              Department of Physics, Astronomy and Mathematics, University of Hertfordshire, Hertfordshire AL10 9AB, UK 
              \and
              Space Science Data Center (SSDC) - Agenzia Spaziale Italiana (ASI), 00133 Roma, Italy
              \and
              University of Messina, Mathematics, Informatics, Physics and Earth Science Department, Via F.S. D’Alcontres 31, Polo Papardo, 98166, Messina, Italy
              \and
              Department of Astronomy, University of Cape Town, Private Bag X3, Rondebosch, 7701, South Africa
              \and
              South African Astronomical Observatory, P.O. Box 9, Observatory, 7935, South Africa
              \and
              The Inter-University Institute for Data Intensive Astronomy, University of Cape Town, Private Bag X3, Rondebosch, 7701, South Africa
              \and
              Centre for Astrophysics and Cosmology, Science Institute, University of Iceland, Dunhagi 5, 107, Reykjavik, Iceland
              \and
              Yunnan Observatories, Chinese Academy of Sciences, Kunming 650011, China
              \and
              INAF – Osservatorio Astronomico di Roma, Via Frascati 33, 00078, Monte Porzio Catone (RM), Italy
              \and
              Anton Pannekoek Institute for Astronomy, University of Amsterdam, P.O. Box 94249, 1090 GE Amsterdam, The Netherlands
              \and
              INAF – Osservatorio di Astrofisica e Scienza dello Spazio, Via Piero Gobetti 93/3, 40129 Bologna, Italy
              \and
              INAF – IASF Milano, Via Alfonso Corti 12, 20133 Milano, Italy
              \and
              Aix Marseille University, CNRS, CNES, LAM, Marseille, France
              \and
              Massachusetts Institute of Technology, Kavli Institute for Astrophysics and Space Research, Cambridge, Massachusetts, United States
              \and
              School of Physics and Astronomy, University of Leicester, University Road, Leicester, LE1 7RH, UK
              \and
              School of Mathematical and Physical Sciences, Macquarie University, NSW 2109, Australia
             }

   \date{Received xxx; accepted yyy}

 
  \abstract
  {Gamma-ray bursts (GRBs) are extremely bright phenomena powered by relativistic jets arising from explosive events at cosmological distances. The nature of the jet and the configuration of the local magnetic fields are still unclear, with the distinction between different models possibly provided by the detection of early-time polarisation.}
  {Past observations do not agree on a universal scenario describing early-time polarisation in GRB afterglows, and new studies are necessary to investigate this open question. We present here the discovery of GRB\,240419A, its redshift determination of $z=5.178$, its early-time optical polarimetry observations, and the multi-wavelength monitoring of its afterglow.}
  {We analysed three epochs of polarimetric data to derive the early-time evolution of the polarisation. The multi-wavelength light curve from the X-rays to the near-infrared band was also investigated to give a broader perspective on the whole event.}
  {We find a high level of polarisation, $P=6.97^{+1.84}_{-1.52}$\,\%, at 1740~s after the GRB trigger, followed by a slight decrease up to $P=4.81^{+1.87}_{-1.53}$\,\% at 3059~s. On the same timescale, the polarisation position angle is nearly constant. The multi-band afterglow at the time of the polarisation measurements is consistent with a forward shock (FS), while the earlier evolution at $t-t_0\lesssim700$~s can be associated with the interplay between the forward and the reverse shocks or with energy injection.}
  {The detected polarised radiation when the afterglow is FS-dominated and the stable position angle are consistent with an ordered magnetic field plus a turbulent component driven by large-scale magnetohydrodynamic instabilities. The lack of a jet break in the light curve prevents a comparison of the polarisation temporal evolution with theoretical expectations from magnetic fields amplified by microscopic-scale turbulence, limiting our ability to constrain the observer's viewing angle. Notably, GRB\,240419A is the most distant GRB with a detected polarised optical afterglow, extending the redshift range for such measurements.}


   \maketitle

\section{Introduction} \label{sec:1}
Gamma-ray bursts (GRBs) are luminous $\gamma$-ray flashes produced by either the collapse of massive stars or the merger of compact binaries. Following this destructive event, a relativistic jet is launched. The observed gamma-ray radiation is produced as electrons are accelerated within the collimated outflow by internal shocks \citep[e.g.][]{Rees&Meszaros92,Meszaros&Rees93,Kobayashi+97,Sari&Piran97,Daigne&Mochkovitch98,Kumar&Piran00} or magnetic reconnection in magnetised outflows \citep{Spruit+01,Drenkhahn02,Drenkhahn&Spruit02,Zhang&Yan11}. This prompt phase typically lasts from fractions up to hundreds of seconds, and the measure of its duration led to the first classification into long and short GRBs, with the separation at around 2 seconds \citep{Kouveliotou+93}. The former have been confirmed to be typically associated with type-Ic broad line (BL)  supernovae by several joint observations \citep[see e.g.][]{Galama+98,Hjorth+03,Fruchter+06,Woosley&Bloom06,Hjorth&Bloom12,Modjaz+16,Cano+17}, whereas the latter arise from a compact binary merger, as has been confirmed by the joint detection of the outstanding gravitational wave event GW170817, the related short GRB\,170817A, and the kilonova (KN) AT2017gfo \citep{Abbott+17,Pian+17,Tanvir+17}. However, this classification is currently challenged by the observations of peculiar events, such as short-duration GRBs with a supernova association \citep[GRB\,200826A][]{Rossi+22}, and long-duration bursts with a clear detection of a related KN (GRB\,211211A, \citealt{Rastinejad+22,Troja+22}, and GRB\,230307A, \citealt{Levan+24}). After the prompt phase, the expanding jet interacts with the surrounding medium and two shocks are produced: a forward shock (FS) powering the afterglow emission, which covers the electromagnetic spectrum from the X-rays up to the radio with different timescales and energetics, and a reverse shock (RS), propagating backwards into the decelerating jet \citep{Piran99,Zhang+03,Zhang&Kobayashi05}. The FS allows us to investigate the energetics, the geometry, and the surrounding environment, while the short-lived RS, whose presence is typically revealed by an optical flash and/or a radio flare, can probe the jet composition and magnetisation \citep[see e.g.][]{Sari&Piran99,Zhang+03,McMahon+06}. \\
\indent More than 20 years of GRB observations with high-energy satellites and ground-based telescopes has revealed a great diversity of events, sometimes challenging the above `fireball' model \citep{Piran99} and highlighting the need for complementary techniques to investigate these events. Polarisation analysis is especially powerful because it provides insights into the jet geometry, the magnetic field configurations in the emitting region, and the outflow composition \citep[for a review, see][]{Covino&Gotz16}. Some degree of polarisation (a few percent) is expected to emerge in the optical flux from the FS as a signature of synchrotron radiation \citep{Meszaros&Rees97}, depending on the morphology and intensity of the shock-generated magnetic field \citep[see e.g.][]{Gruzinov&Waxman99,Medvedev&Loeb99}. The magnetic field structure, together with the observer's viewing angle and the jet composition, strongly impacts the polarisation degree, $P$, which is expected to evolve with time. In particular, the magnetic fields can originate and be amplified through either kinetic or magnetohydrodynamic (MHD) instabilities. Kinetic processes such as the Weibel instability generate small-scale, tangled fields on plasma skin depth scales, whereas large-scale, turbulent fields on blast wave thickness scales can result from MHD instabilities. These two regimes produce distinct polarisation signatures, particularly during the early afterglow \citep[see][]{Rossi+04,Teboul&Shaviv21,Kuwata+23,Kuwata+24}. In addition, geometrical models linking the variation in the FS polarisation degree and position angle to the afterglow light curve evolution \citep{Ghisellini&Lazzati99,Sari99} have been successfully deployed to explain afterglow polarimetric observations in some bursts, including GRB\,021004 \citep{Lazzati+03,Rol+03}, GRB\,091018 \citep{Wiersema+12}, and GRB\,121024A \citep{Wiersema+14}. A very bright source and a large signal-to-noise ratio (S/N) in the observations are also needed for successful polarisation detection. In the ideal case of polarisation detection at multiple epochs, although uncommon, it is possible to infer the intrinsic parameters of the source. If such detections are secured before and after the jet break in the afterglow light curve, i.e. the time when the broad-band emission undergoes an achromatic steepening \citep{Rhoads99}, its diagnostic power is even more powerful. Such cases can reveal specific features in the polarisation curve \citep[e.g. a 90$^\circ$ position angle rotation,][]{Ghisellini&Lazzati99}, and, more generally, the polarisation time evolution also depends on the jet break time \citep{Rossi+04}. \\
\indent The RS can be highly polarised if an ordered magnetic field advected from the central source is present \citep{Granot&Konigl03,Lyutikov03}. In particular, we expect $P$ ranging from $\sim10$\% to $\sim60$\% in the optical band in case of tangled or large-scale ordered fields, respectively. This has a relevant diagnostic power, since it can provide a measure of the jet magnetisation parameter, $\sigma_B$ \citep[as for GRB\,120308A, see][]{Mundell+13,Zhang+15}, which impacts the relative luminosity of the RS compared to that of the FS. In the case of a polarised RS, we also expect a stable or a randomly varying polarisation position angle depending on the specific magnetic field configuration and instabilities present in the emitting region \citep{Gruzinov&Waxman99}. Hereafter, we only discuss linear polarisation, which we refer to just as polarisation. \\
\indent Extensive polarimetric observations of late-time optical afterglows have been obtained in the past years, showing a polarisation degree at a typical level of a few percent \citep[e.g.][]{Covino+99,Rol+00,Covino+03,Greiner+03,Gorosabel+04,Wiersema+12,Wiersema+14,Urata+23,Brivio+22,Agui+24}. This was usually associated with the FS emission from the shocked ambient medium, which dominates at these times, with low polarised radiation expected. On the other hand, at earlier times (before 10\,000 seconds from the GRB trigger), some optical afterglows show a level of polarisation up to tens of percent (see Table~\ref{Tab:pol_examples}). This larger polarised emission has been associated with the presence of ordered magnetic fields within the jet. In particular, a high level of polarisation at an early time may be the signature of a RS, even if studies on previous observations do not always agree on this point \citep{Steele+09,Uehara+12,Mundell+13}. Thus, additional early polarimetric observations are crucial to distinguish among different origins; for example, a pure FS \citep{Uehara+12,Urata+23,Mandarakas+23}, a dominant RS \citep{Steele+09,Mundell+13,Steele+17}, or cases showing polarisation in both FS and RS regimes \citep{Arimoto+24}, with additional mechanisms (e.g. energy injection, \citealt{Shrestha+22}; or refreshed shocks, \citealt{Agui+24}) possibly contributing. Additional observations of early-time optical polarisation have been obtained for other events during the prompt phase. For example, GRB\,160625B exhibited a relatively high polarisation degree (up to $\sim8.3$\,\%) within 330 seconds from the burst trigger, although its origin is likely contaminated by the prompt emission \citep{Troja+17}. A similar interpretation was provided for the early time polarisation ($P=7.7\pm1.1$\,\%) accompanying GRB\,190114C \citep{Jordana+20}. This makes a direct comparison with polarisation arising from the afterglow phase difficult, and we exclude such cases from further considerations. \\
\indent In this work, we present the discovery of GRB\,240419A, detected by the {\it Neil Gehrels Swift Observatory} \citep[hereafter {\it Swift},][]{Gehrels+04}, and the analysis of its afterglow. Polarisation measurements from around 30 minutes from the burst trigger were secured, yielding a high level of optical polarisation. Additional optical and near-infrared (NIR) imaging observations were obtained at comparable timescales. The spectral analysis revealed a remarkable redshift of $z=5.178$ \citep{GCN_UVES}, making it the most distant GRB with polarised optical afterglow ever detected. This work is organised as follows: in Sect.~\ref{sec:2}, we present the observations carried out for GRB\,240419A, while in Sect.~\ref{sec:3} we describe the data analysis and present the results for the polarisation data and the broadband afterglow. In Sect.~\ref{sec:4}, we provide a detailed discussion of these results, while our conclusions are summarised in Sect.~\ref{sec:5}.\\
We adopt the $\Lambda$CDM model with cosmological parameters $\Omega_M=0.308$, $\Omega_\Lambda=0.692$, and $H_0=67.8$ km~s$^{-1}$~Mpc$^{-1}$ \citep{PLANCK16}. All magnitudes presented in this work are given in the AB system. Unless otherwise stated, errors are at the $1\sigma$ confidence level (c.l.) and upper limits at a $3\sigma$. For the flux density of the afterglow, we adopt the convention $F_\nu(t) \propto t^{\alpha}\nu^{\beta}$.

\begin{table*}[ht]
    \centering
    \caption{GRBs with detected early-time ($t-t_0\lesssim10\,000$~s) polarised optical light from the afterglow phase.}
    \footnotesize
    \renewcommand{\arraystretch}{1.1}
    \begin{tabular}{lccccl}
    \hline
    \noalign{\smallskip}
    GRB & $z$& $P$ [\%]& $t-t_0$ [s]& Rest frame $t-t_0$ [s]& Reference\\
    \noalign{\smallskip}
    \hline
    \hline
    \noalign{\smallskip}
    090102 & 1.55 & $10.2\pm1.3$ & 190 & 74.5 & \cite{Steele+09}\\
    091208B & 1.063 & $10.4\pm2.5$ & 149--706 & 72--342 & \cite{Uehara+12} \\
    101112A & $\lesssim3.5$ & $6^{+3}_{-2}$ & 176--355 & -- & \cite{Steele+17} \\
    & & $6^{+4}_{-3}$ & 715--893 & -- & \\
    110205A & 2.22 & $13^{+13}_{-9}$ & 240--840 & 74.5--291 & \cite{Steele+17} \\
    120308A & 2.2 & $28^{+4}_{-4}$ & 240 & 75 & \cite{Mundell+13} \\
    & & $16^{+5}_{-4}$ & 800 & 250 & \\
    121024A & 2.298 & $4.09\pm0.20$ & 9698 & 2854 & \cite{Wiersema+14} \\
    180720B & 0.654 & $\sim0.5-8$ & 80--2000 & 48.4--1209 & \cite{Arimoto+24} \\
    190114C & 0.4245 & $7.7\pm1.1$ & 52 & 36.5 & \cite{Jordana+20} \\
    & & $\sim2-4$ & 108--2000 & 76--1375& \\
    191016A & 3.3 & $4.7\pm4.1$& 3987--4587& 927--1067 & \cite{Shrestha+22} \\
    & & $11.2\pm6.6$& 4587--5187& 1067--1206 & \\
    & & $14.6\pm7.2$& 5787--6387& 1346--1485 & \\
    191221B & 1.148 & $1.4\pm0.1$ & 10454 & 4867 & \cite{Urata+23} \\
    210610B & 1.1341 & $4.27\pm1.45$ & 10411 & 4878.5 & \cite{Agui+24} \\
    210619B & 1.937 & $2.2\pm0.7$ & 6537 & 2225.7 & \cite{Mandarakas+23} \\
    & & $2.6\pm0.8$ & 6751 & 2298.6 & \\
    \noalign{\smallskip}
    \hline
    \end{tabular}
    \label{Tab:pol_examples}
\end{table*}

\section{Observations} \label{sec:2}
\subsection{Gamma and X-rays}\label{sec:2-Swift}
GRB\,240419A was detected by the {\it Swift} Burst Alert Telescope \citep[BAT;][]{BAT} on April 19 2024 at $t_0=$ 01:48:01 UT \citep{GCN_Swift}, at a refined position, (J2000) R.A. = 06:18:01.3, Dec. = --44:57:50.2, with an uncertainty of $3.0^{\prime}$ \citep[radius, 90\% c.l.,][]{GCN_BATpos}. The peak count rate is $\sim1500$~counts~s$^{-1}$ (in the 15--350 keV band) at $\sim1$~s after the trigger \citep{GCN_BATpos}. The measured duration of the burst is $T_{90}=114.26\pm33.13$~s. The fluence in the 15--150 keV band is $6.3\pm1.3\times10^{-7}$ erg~cm$^{-2}$ and the 1-s peak photon flux measured from $t_0+0.35$~s in the same band is $0.7\pm0.1$ ph~cm$^{-2}$~s$^{-1}$. The best fit for the time-averaged spectrum is provided by a simple power-law model, with photon index $\Gamma=1.64\pm0.32$ (errors at 90\% c.l.).\\
\indent The {\it Swift} X-ray Telescope \citep[XRT;][]{XRT} began observing the field at 01:49:51.1 UT; that is, 110.1~s after the BAT trigger, and found a bright, uncatalogued, fading X-ray source, located at the enhanced position, (J2000) R.A. = 06:18:08.17, Dec. = --44:59:58.3, with an uncertainty of $2.2^{\prime\prime}$ \citep[radius, 90\% c.l.,][]{GCN_XRTpos}. Follow-up observations were carried out up to $\sim1.9$~d after $t_0$, when the source became too faint to be detected. The analysis of the {\it Swift}/XRT observations is presented in Sect.~\ref{sec:3-XRT}.

\subsection{Optical/NIR afterglow}\label{sec:2-opt/NIR}
\begin{figure}
    \centering
    \includegraphics[width=0.45\textwidth]{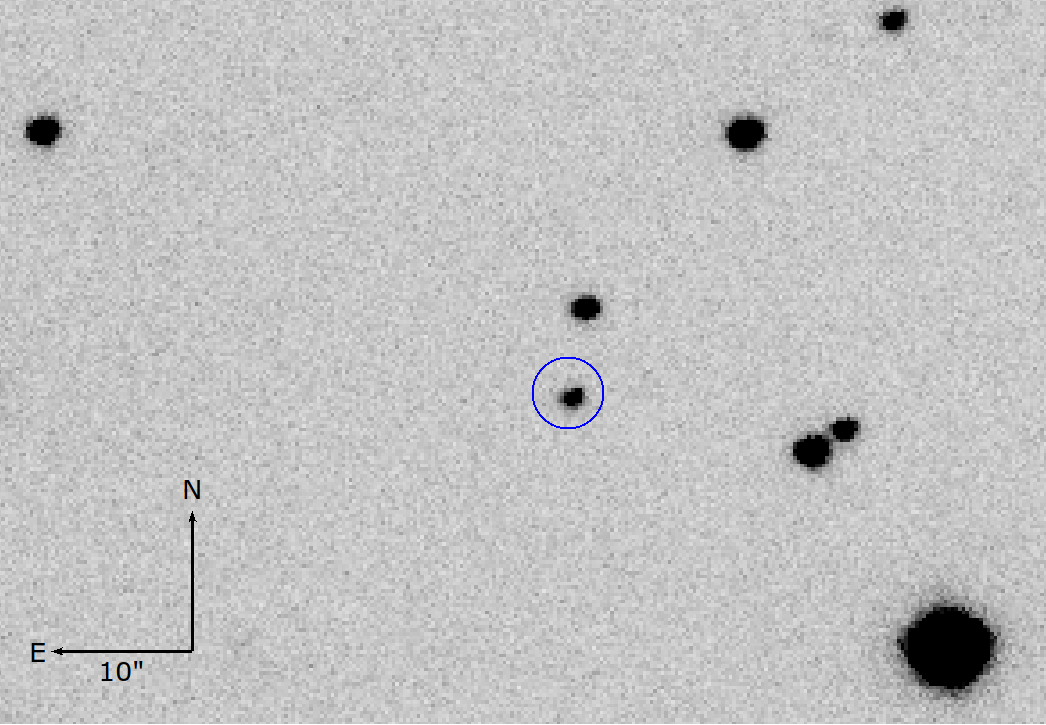}
    \caption{Acquisition image of the field of GRB\,240419A observed by VLT/FORS2 on April 19 2024 at $t-t_0\sim20.9$~min. The blue circle marks the position of the optical afterglow.}
    \label{fig:fc}
\end{figure}

The field of GRB\,240419A was observed from the ground with optical and NIR telescopes starting just a few minutes after the trigger. We present here all the observations we carried out and all the results of the observations described in the following are reported in Table~\ref{Tab:photometry}.\\
\indent The optical/NIR afterglow was discovered by the 0.6m robotic Rapid Eye Mount telescope \citep[REM;][]{REM_Zerbi01,REM_Covino+04}, located at the European Southern Observatory (ESO) in La Silla (Chile). The observations started on April 19 2024 at 01:49.38 UT, 97 s after the burst, and lasted for about 3 hours. All the images were automatically reduced through the {\tt jitter} tool of the {\tt eclipse} package \citep{eclipse}, which performs image registration, sky estimation, and subtraction through the median combination of five individual images to obtain one average frame for each sequence. A variable source was found within the XRT error circle in all $H$-band images \citep{GCN_REM} and from the early-time observations in the $i$ band. Upper limits were derived for the $g$ and $r$ optical bands, while the presence of strong fringes in the $z$-band frames did not allow us to perform a reliable analysis on such images. The $J$ and $K$ filters were not available at the time of the observations. \\
\indent The optical afterglow was also observed with the FOcal Reducer/low dispersion Spectrograph 2 \citep[FORS2;][]{FORS2} mounted on the ESO Very Large Telescope (VLT) Unit Telescope 1 (UT1-Antu), at Cerro Paranal, Chile. The observations began automatically in rapid response mode (RRM)\footnote{ESO designed the RRM protocol to automatically trigger observations of GRBs by interrupting ongoing observations and quickly redirecting telescopes towards the target in about 8-10 minutes.} on April 19 2024, at $t-t_0\sim20.9$~min. Observations consist of three images in the $R_{\rm special}$ filter ($\lambda_0=6482$ \AA, ${\rm FWHM}=1645.55$ \AA) at the relatively high airmass of 2.1, with an exposure time of 20 s each. Data reduction was carried out following the standard procedures: after bias subtraction, non-uniformities were corrected using a normalised flat-field frame processed with tools from the Swift Reduction Package (SRP)\footnote{\url{http://www.me.oa-brera.inaf.it/utenti/covino/usermanual.html}}. The afterglow was detected within the XRT error circle at the coordinates (J2000): R.A. = 06:18:08.17, Dec = $-$44:59:57.3 \citep{GCN_FORS} with an uncertainty of 0.5$^{\prime\prime}$. The astrometry of the field was determined using the SkyMapper DR4 \citep{Skymapper_DR4} catalogue. An acquisition image displaying the position of the optical afterglow is shown in Fig.~\ref{fig:fc}. FORS2 then observed the field in imaging polarimetry (IPOL) mode for $\sim30$ min, from $\sim26.6$ to $58.0$~min after the trigger. The target was detected in all frames and the details about the polarisation analysis are reported in Sect.~\ref{sec:3-Pol}. The {\it Swift} UltraViolet-Optical Telescope \citep[UVOT;][]{UVOT} also observed the field but found no source consistent with the XRT position, and upper limits were derived \citep{GCN_UVOT}.\\
\indent Further observations of GRB\,240419A were also obtained with the ESO-VLT UT2 (Kueyen) equipped with the Ultraviolet and Visual Echelle Spectrograph \citep[UVES;][]{UVES}, still in the RRM mode \citep{GCN_UVES}. Observations started at 02:15:42 UT ($27.7$~min after the GRB) at an average airmass of 2.4 and the spectra covered a wavelength range from 3700 to $\sim9400$ \AA, with a total exposure time of 30 min. The analysis of the spectrum is presented in Sect.~\ref{sec:3-spec}.\\
\indent We obtained imaging of the field of GRB\,240419A with the 0.6~m wide-field MeerLICHT optical telescope \citep{MeerLICHT} located at the South African Astronomical Observatory (SAAO) site in Sutherland, South Africa. Observations began at $t-t_0\sim0.65$~d and consisted of a series of 300~s exposures in the $i$, $z$, and $q$ (440–720 nm) filters. No optical source was detected at the position of GRB 240419A in any of the individual exposures. We also performed observations with the BlackGEM telescope array \citep{BlackGEM} located at the ESO La Silla Observatory in Chile. Observations were obtained with unit telescope 2 (BG2) and consisted of $2\times300$~s exposures in each of the $i$ and $z$ bands, with the first exposure starting at $\sim0.89$~d post-trigger. We detected the optical afterglow in a single $z$-band frame but not in the other images.  

\subsection{Other observations}
The event was followed up with the Australia Telescope Compact Array (ATCA) at 5.5, 9, and 18 GHz as part of the Long-term ATCA `PanRadio GRB' follow-up programme C3542 (PI Anderson). ATCA began observing at 04:00 UT on April 19 2024 (132 min post-burst) up until 12:30 UT. A source coincident with the XRT position was marginally detected with $63\pm18~\mu$Jy/beam at 18 GHz \citep{GCN_ATCA}. Upper limits at 39 and 42 $\mu$Jy/beam were obtained at 5.5 and 9 GHz, respectively.

\section{Data analysis and results} \label{sec:3}

\subsection{X-rays}\label{sec:3-XRT}
\begin{figure}
    \centering
    \includegraphics[width=.45\textwidth]{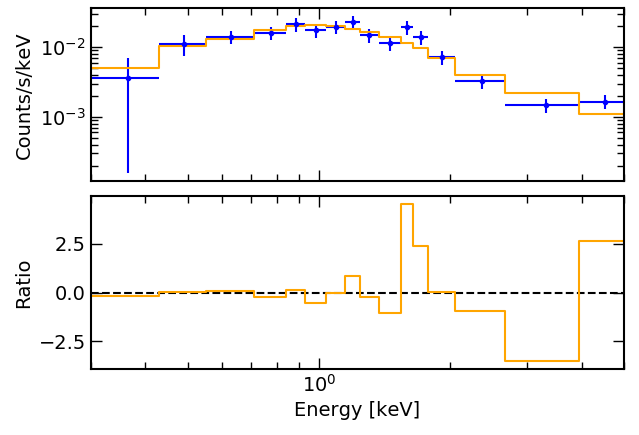}
    \caption{{\it Swift}/XRT PC mode spectrum of GRB\,240419A. The best fit is shown in orange (top panel). The ratio between the data and the folded model is displayed in the bottom panel.}
    \label{fig:XRT_spec}
\end{figure}

We retrieved the count-rate {\it Swift}/XRT light curve of GRB\,240419A from the GRB XRT light curve repository\footnote{\url{https://www.swift.ac.uk/xrt\_curves/01222955}} \citep{Evans+07,Evans+09}. The light curve comprises 19.9 ks of data obtained between $t_0+117$ s and $t_0+162.2$ ks, all in photon counting (PC) mode. The light curve can be fitted with a double broken power law. It initially decays with an index $\alpha_{X,1}=-5.4\pm0.3$, then it follows a slope $\alpha_{X,2}=0.1\pm0.2$ from $246\pm15$ s to $1629\pm303$ s after $t_0$, and at later times it decays with $\alpha_{X,3}=-1.2\pm0.1$. The resulting fit statistics is $\chi^2{\rm /d.o.f.}=4.34/10$. We display the X-ray light curve, along with its best fit (dashed blue line), in Fig.~\ref{fig:lightcurve}.\\
\indent We also retrieved spectral data from the {\it Swift}/XRT GRB spectrum repository\footnote{\url{https://www.swift.ac.uk/xrt_spectra/01222955}} and analysed the spectrum. The source spectrum was first grouped to have at least 20 counts per bin and fitted with an absorbed power-law model within the {\tt XSpec} package \citep[Version 12.14.1,][]{Xspec}, keeping the Galactic contribution $N_{\rm H}$ and the redshift fixed at $6.63\times10^{20}$~cm$^{-2}$ \citep{Willingale+13} and 5.178 (see Sect.~\ref{sec:3-spec}), respectively. We obtained a photon spectral index of $\Gamma=2.17^{+0.18}_{-0.17}$ and a best-fitting intrinsic column density of $N_{{\rm H},z}=1.0^{+0.6}_{-0.3}\times10^{23}$~cm$^{-2}$, in excess of the Galactic value. The resulting fit statistics is $\chi^2{\rm /d.o.f.}=15.01/16$. The spectrum and its best fit are shown in Fig.~\ref{fig:XRT_spec}.

\subsection{Spectral analysis}\label{sec:3-spec}
The optical spectrum of GRB\,240419A was obtained in RRM mode with UVES. The UVES spectrograph uses a dichroic system to split incoming light into two separate arms (BLUE and RED, with a wavelength coverage of 3000--5000\,\AA\ and 4200--11000\,\AA, respectively), allowing for simultaneous observation across a wide wavelength range with high resolving power ($\sim40,000$). There are instrumental gaps between the various arms and the different dichroics used. The full dataset was reduced using the ESO pipeline \citep{Esoreflex} once the final calibration has been released. The observation had to be cut short as the telescope reached the elevation limit of 20 degrees. Hence, the full RRM observation was not completed (only 20 minutes with respect to the planned RRM observation block).\\
\indent The final spectrum was produced by weighting by the exposure time and stacking the single pieces. Two standard stars, LTT\,3218 and LTT\,386, were observed for flux calibration $\sim3$ hours after the spectral observations, at an average airmass of 1.1. The flux calibration was carried out following the procedure described in the ESO instrument's user manual\footnote{\url{https://ftp.eso.org/pub/dfs/pipelines/instruments/uves/uves-pipeline-manual-6.5.3.pdf}}. Wavelengths were corrected to the vacuum-heliocentric system. Inspecting the entire spectrum, the trace signal is clearly visible in the wavelength range from 7700 \AA\, to $\sim9400$ \AA, while the bluer side has zero flux. We detected several absorption lines from both low- and high-ionisation transitions as being due to \sii, \siii, \oi, \cii, and \siiv. Fine-structure lines such as \siiis and \ciis were also identified. From all the identified absorption lines\footnote{\sii \,$\lambda$1250 , \sii \,$\lambda$1253 , \sii \,$\lambda$1259 , \siii \,$\lambda$1260 , \siii \,$\lambda$1304 , \oi \, $\lambda$1302 , \cii \, $\lambda$1334 , \siiv \, $\lambda$1393 , \siiv \,$\lambda$1402 , \siiis \, $\lambda$1264 , \ciis \, $\lambda$1335 } we infer a redshift of $z_{\rm GRB}= 5.1777\pm0.0002$. The host galaxy system spans $\sim350$\,km\,s$^{-1}$.\\
\indent From low-ionisation transitions, there are clearly two components (with $\Delta v\sim\,$210\,km\,s$^{-1}$) at $z=5.1777\pm0.0002$ (the strongest and main component\footnote{The uncertainties on the redshift measurements represent the standard deviation of the redshift values obtained from the individual lines identified. Please note the presence of multiple components: we assume the redshift of the strongest component as $z_{\rm GRB}$ with its uncertainty.}) and $z=5.1734\pm0.0002$, as shown in Fig. \ref{fig:spec}. The high-ionisation transitions (\siiv$\lambda$1393 \AA, $\lambda$1402 \AA) show as well two components but blueshifted ($\sim\,-$30\,km\,s$^{-1}$) compared to the low-ionisation lines. The Lyman-$\alpha$ transition falls in a region of the spectrum where an instrumental gap is present. As a result, it was not possible to fit the absorption due to the neutral hydrogen and thus infer $\log (N$(H\,{\sc i})/cm$^{-2}$). Moreover, the limited S/N (average $\sim1.2$) prevented us from performing a Voigt profile fit and determining the column densities.

\subsection{Polarisation analysis}\label{sec:3-Pol}
\subsubsection{Observations and data analysis}
The polarimetric observations of GRB\,240419A were obtained in the IPOL mode with the $R_{\rm special}$ filter. The dataset includes three observing cycles, each of them with four exposures at four different angles (0$^{\circ}$, 22.5$^{\circ}$, 45$^{\circ}$, 67.5$^{\circ}$) of the half-wave plate in the instrumental setting of FORS2 (see Table~\ref{Tab:Pol_log}). The IPOL mode ensures the acquisition of such frames through the use of a Wollaston prism splitting the image of each object in the field into the two orthogonal polarisation components appearing in adjacent areas of the image. A strip mask is used in the focal area of the instrument to avoid the overlap of the two beams of polarised light on the CCD. In this way, for each position angle $\phi/2$ of the half-wave plate rotator, we obtain two simultaneous images of cross-polarisation at angles $\phi$ and $\phi+90^{\circ}$. In addition to the optical afterglow, observations of two polarised standard stars, Hiltner\,652 and BD-12\,5133, were considered to fix the offset between the measured polarisation angle and the instrumental reference frame. Data reduction was carried out in a standard manner: after bias subtraction, non-uniformities were corrected using flat fields obtained without the Wollaston prism \citep[see e.g.][]{Patat&Romaniello06}. We performed aperture and point-spread function (PSF) photometry to measure the flux of the point sources in the field with the DAOPHOT \citep{Daophot_Stetson87} and ALLSTAR packages. Apertures were chosen to be $\sim1.5$ times the full width at half maximum (FWHM), measured individually for each point source in every image. Sky subtraction was performed using annuli of inner and outer radii of four and five times the FWHM. Each pair of simultaneous measurements at orthogonal angles was used to compute the $Q/I, U/I$ reduced Stokes parameters. This technique removes any difference between the two optical paths (ordinary and extraordinary rays) and minimises the impact of polarimetric flat-fielding errors. Moreover, being based on relative photometry in simultaneous images, our measurements are insensitive to intrinsic variations in the optical transient flux. In addition, some bright, nearby field stars were investigated to look for possible spurious contributions, because stars are typically unpolarised sources, except for some polarisation induced by the dust along the sightline towards the star itself. The reduced Stokes parameters $Q/I$ and $U/I$ describing the linear polarisation of the radiation were derived using the following formulae:
\begin{eqnarray}
    \begin{aligned}
    \frac{Q}{I} &= \frac{1}{2}\bigg(\frac{f_o-f_e}{f_o+f_e}\bigg|_{0^{\circ}}-\frac{f_o-f_e}{f_o+f_e}\bigg|_{45^{\circ}}\bigg)\\
    \frac{U}{I} &= \frac{1}{2}\bigg(\frac{f_o-f_e}{f_o+f_e}\bigg|_{22.5^{\circ}}-\frac{f_o-f_e}{f_o+f_e}\bigg|_{67.5^{\circ}}\bigg),
    \end{aligned}
    \label{eq:qu}
\end{eqnarray}
which include all data obtained at the four angles of the half-wave plate. The subscripts $_o$ and $_e$ represent the ordinary and extraordinary rays, respectively, into which the incoming radiation was split. Errors on the reduced Stokes parameters were computed via standard propagation theory.

\subsubsection{Results}\label{sec:3-pol_results}

\begin{figure*}[!ht]
    \centering
    \includegraphics[width=.85\textwidth]{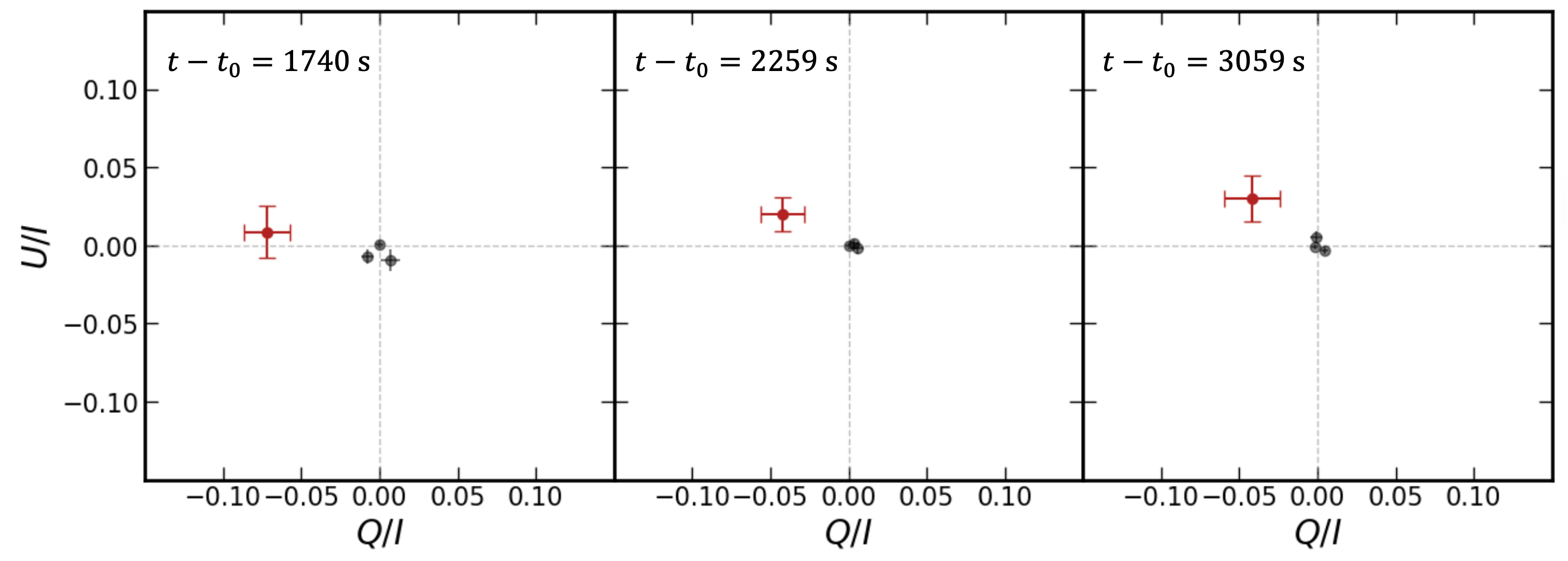}
    \caption{$Q,U$ plots for the three imaging polarimetric observations. Error bars represent $1\sigma$ uncertainties. The afterglow is marked with the red bars and shows significant polarisation with respect to the (unpolarised) field stars (black dots).}
    \label{fig:QU_plots}
\end{figure*}

\begin{table}[!ht]
    \caption{Results of GRB\,240419A optical afterglow polarimetric observations.}
    \centering
    \tiny
    \renewcommand{\arraystretch}{1.3}
    \begin{tabular}{ccccc}
    \hline
    \noalign{\smallskip}
    $t - t_0$ [s]& $Q/I$& $U/I$& $P$ [\%]& $\theta$ [$^{\circ}$]\\
    \noalign{\smallskip}
    \hline
    \hline
    \noalign{\smallskip}
    1740& $-$0.071 $\pm$ 0.016& 0.008 $\pm$ 0.015& 6.971$^{+1.835}_{-1.519}$& 86.98 $\pm$ 5.92\\
    2259& $-$0.041$\pm$ 0.011& 0.023$\pm$ 0.014& 4.577$^{+1.306}_{-1.083}$& 75.61 $\pm$ 8.04\\
    3059& $-$0.039 $\pm$ 0.015& 0.032$\pm$ 0.018& 4.813$^{+1.868}_{-1.529}$& 70.20 $\pm$ 9.50\\
    \noalign{\smallskip}
    \hline
    \end{tabular}
\tablefoot{Summary of the results for imaging polarimetry frames analysis. Errors on the reduced Stokes parameters and the position angle were computed via propagation theory; those on $P$ were obtained after bias correction when appropriate (see Sect.~\ref{sec:3-Pol}). Uncertainties are at $1\sigma$ level.}
    \label{Tab:result}
\end{table}

To obtain the polarisation degree associated with the optical afterglow, we first derived the polarisation of three field stars to subtract their contribution to the afterglow polarisation: all of them are isolated, unsaturated in every epoch, and at least comparable in brightness to the afterglow. They were also selected to lie within 1$^\prime$ of the optical afterglow position to minimise the instrumental polarisation induced by the distance from the optical axis. The correction, applied using the corresponding $Q$, $U$ background values and the polarisation maps presented by \cite{Gonzalez+20}, results in a residual instrumental polarisation of the order of $\sim$0.01\,\%. The average contribution given by the field stars is: $\langle Q/I \rangle=0.0001\pm 0.0038, \langle U/I \rangle=-0.0014\pm0.0026$ in epoch 1, $\langle Q/I \rangle=0.0016\pm0.0018, \langle U/I \rangle=0.0016\pm0.0005$ in epoch 2, and $\langle Q/I \rangle=0.0009\pm0.0018, \langle U/I \rangle=0.0001\pm0.0025$ in epoch 3. Therefore, the polarisation induced by the local interstellar matter is low, and its contribution to the observed polarisation can be considered negligible. We also estimated the expected contribution of the interstellar polarisation from the Galactic extinction and found a negligible effect $<P_{\rm ISP}>=3.5\%\times E(B-V)^{0.8}=0.32\,\%$ \citep{Fosalba+02}, as expected from the average polarisation of the field stars. Then, we analysed the polarised standard star Hiltner\,652 in order to check for possible polarisation angle correction: we derived $\theta_{\rm H}=179.74\pm0.82$ deg, while the expected one is $\theta_{\rm exp,H}=179.39\pm0.03$ deg \citep{Cikota+17}. Being consistent at the $1\sigma$ level, no correction is needed. The other polarised standard star observed, BD-12\,5133, was saturated in our observations, making its analysis unreliable.\\
\indent We thus derived the reduced $Q/I, U/I$ Stokes parameters associated with the optical afterglow of GRB\,240419A. We show the results in the $Q,U$ plots in Fig.~\ref{fig:QU_plots}: the optical afterglow displays significant polarisation despite the large errors. The degree, $P$, and angle, $\theta$, of polarisation were obtained from the measurements of $Q/I$ and $U/I$: $P=\sqrt{Q^2+U^2}/I$, $\theta=\frac{1}{2}\arctan(U/Q)$. Moreover, for any low level of polarisation, ($P/\sigma_P\leq4$), the distribution function of $P$ and of $\theta$ is no longer normal and that of $P$ becomes skewed (following a Rice distribution, \citealt{Rice44}; see also \citealt{Patat&Romaniello06}). A correction taking into account this bias is required, and we adopted the modified asymptotic estimator defined by \cite{Plaszczynski+14} to derive the correct value of the polarisation degree, $P$. A similar correction for the polarisation position angle is usually negligible \citep{Montier+15} and therefore it was not applied in this case. We report the final results in Table~\ref{Tab:result}.\\
\indent Our polarimetric observations could, in principle, be biased by the use of the $R_{\rm special}$ filter, which lies within the Lyman-$\alpha$ forest at the GRB redshift. However, while dust within neutral hydrogen clouds in the intergalactic medium may produce scattering-induced polarisation \citep[e.g.][]{Humphrey+13}, the Lyman-$\alpha$ forest consists of numerous, randomly distributed absorption systems along the line of sight. As a result, any dust grains within these systems are unlikely to share a coherent alignment, making a significant net induced polarisation improbable. Nevertheless, if a single intervening abosrber were to dominate within the observed band, it could, in principle, produce a detectable additional polarisation component. The variability we detect suggests that the measured polarisation is predominantly intrinsic to the afterglow, although the presence of a constant component from line-of-sight material cannot be fully excluded. Ultimately, polarimetric observations within the Lyman forest and redwards of the Lyman-$\alpha$ break in future GRB afterglows will be required to test this scenario.

\subsection{Light curve analysis}\label{sec:3-lc}

\subsubsection{Temporal and spectral fitting}
\begin{figure*}
    \centering
    \sidecaption
    \includegraphics[width=12cm]{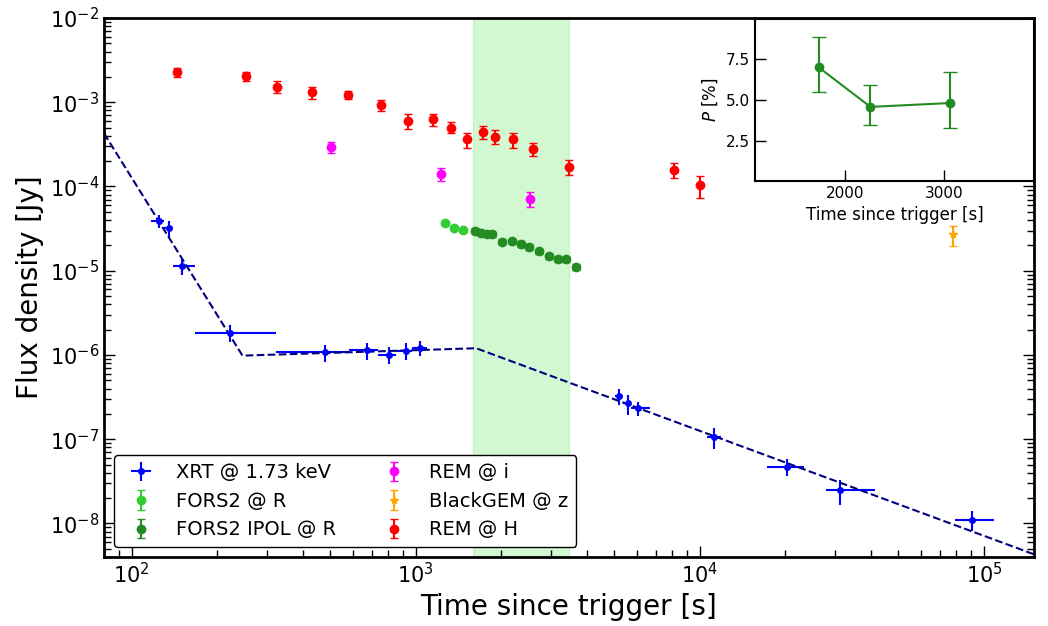}
    \caption[Multi-wavelength light curve of GRB\,240419A]{Multi-wavelength light curve of GRB\,240419A. XRT flux densities were computed at 1.73 keV, the log-mean of the XRT band. The dashed line represents the best fit of the X-ray curve. The shaded light green area marks the time of polarisation observations and the polarisation curve is shown in the top right inset. Upper limits are not shown for display purposes.}
    \label{fig:lightcurve}
\end{figure*}

In addition to the polarisation analysis, we analysed the FORS2 acquisition frames and REM images to derive the optical afterglow light curve. We performed aperture photometry with the \texttt{SExtractor} package \citep[Version 2.28.2][]{SExtractor} and calibrated our results against the SkyMapper DR4 catalogue \citep{Skymapper_DR4}. To derive the $R$-band curve we analysed both the FORS2 acquisition images and the IPOL observations. For every IPOL frame, we summed the fluxes of the optical transient and the reference stars in the ordinary and the extraordinary rays to derive their total intensity. We also performed photometry on the NIR images with the same packages and calibrated them against the 2MASS catalogue \citep{2MASS}. We report all the results in Table~\ref{Tab:photometry} and show them, together with the {\it Swift}/XRT curve, in Fig.~\ref{fig:lightcurve}.\\
\indent We modelled the available optical-NIR photometry with a broken power law (BPL) and a simple power law (PL) model for the temporal and spectral behaviour (of the two bands redder than Ly$_\alpha$), respectively. The modelling was carried out with the {\tt JSPEC} package\footnote{\url{https://github.com/stefanocovino/JSPEC.jl.git}} in a Bayesian framework, with the optical-NIR extinction modelled assuming a Small Magellanic Cloud extinction curve \citep{Gordonetal2016} as implemented in the {\tt Dust Extinction} package\footnote{\url{https://github.com/JuliaAstro/DustExtinction.jl}}. Other extinction recipes (i.e. those modelling Large Magellanic Cloud and Milky Way average environments) did not provide an acceptable fit for the data. We did not consider the $R$-band data in the spectral modelling because they are bluer than the Lyman-$\alpha$ limit considering the GRB redshift and we added some photometric points obtained from the UVES spectrum with a free normalisation, given the priors, to take care of possible uncertainties in the absolute calibration of UVES spectra (best fit value $0.99 \pm 0.31$). We obtained four points just by integrating the flux spectrum in $\sim 100$\AA\ wide bands free from strong absorptions and expressing the results as flux densities. The model was developed within the {\tt Turing} probabilistic framework\footnote{\url{https://turing.ml/}}. Priors were defined as large Gaussian centred on typical results from past GRB modellings for the spectral and temporal slopes, while for extinction, treated as a logarithmic quantity, a uniform prior was chosen. Similarly, uniform priors in log space were used for the temporal break. All adopted priors are listed in Table~\ref{tab:priors}. We verified that the results are independent of the specific choice of (broad) prior parameters by comparing results with different (physically acceptable) prior choices (e.g, changing the Gaussian centres and widths). We also computed a `Bayesian $p_B$-value', following \citet{Lucy16}, to evaluate the fit quality, and found $p_B \sim 0.8$.\\
\indent The optical-NIR light curve decays with index $\alpha_1=-0.52^{+0.13}_{-0.11}$ up to $T_{\rm break} = 691^{+235}_{-147}$~s, then it follows a slope $\alpha_2=-1.07^{+0.05}_{-0.05}$. The spectral index is $\beta=-1.4\pm0.3$, and we obtain an upper limit on the intrinsic optical-NIR extinction, $E(B-V)<0.066$ mag (99\% c.l.). The light curves with their modelling are shown in Fig.~\ref{fig:lc_model}, with the corresponding contour plots presented in Fig.~\ref{fig:lc_contours}. The spectral energy distribution (SED) with the associated model fit is displayed in Fig.~\ref{fig:SED}. The post-break decay slope and the spectral index are consistent with the expectations of synchrotron emission from a GRB FS, according to the GRB closure relations \citep[see e.g.][]{Sari+98,Zhang&Meszaros04}, if the optical band lies above the frequency corresponding to the minimum Lorentz factor of the accelerated electrons ($\nu_m$) and the cooling frequency ($\nu_c$). Moreover, the X-ray spectral index derived in Sect.~\ref{sec:3-XRT} is consistent with $\beta$, and $\alpha_{X,3}$ closely matches optical–NIR decay $\alpha_2$, both agreeing with typical values for GRB afterglows \citep{Zhang+06,Piran99}. These results support a scenario in which the observed emission after $T_{\rm break}$ originated from a pure FS in the regime $\nu_X>\nu_{\rm opt}>{\rm max}(\nu_m,\nu_c)$. From the late-time slope we also estimated the index of the electron distribution ($n(e)\propto \gamma^{-p}$), $p\simeq2.1\pm0.1$, which is consistent with typical GRB afterglow values in the range $2<p<3$ \citep{Panaitescu&Kumar00,Panaitescu&Kumar01,Zhang&Meszaros04}. We also investigated if the break observed in the optical-NIR curve can be interpreted as a jet break. In this scenario, the post-break decay should steepen to $\alpha\sim p\sim 2-3$, which is not consistent with the value derived for $\alpha_2$. In addition, based on the GRB closure relations in the same regime identified above, a jet break would produce a change in the decay slope of $1.0\lesssim\Delta\alpha\lesssim1.25$ for $2<p<3$, significantly larger than the observed $\Delta\alpha = 0.55$. These considerations strongly disfavour the presence of a jet break in the light curve. From the limit derived on the intrinsic extinction, we also estimated the maximum contribution to the observed polarisation from the dust in the host galaxy, assuming a Serkowski \citep{Serkowski+75} law due to the limited information on local dust properties. This yielded an upper limit of $P_{\rm HG} \leq9\,\%\times E(B-V)=0.59\,\%$, indicating a negligible contribution.

\subsubsection{Physical modelling}\label{sec:3-modeling}
We also simultaneously fitted all available multi-wavelength data (X-ray, NIR, and radio observations) using the \texttt{afterglowpy} package \citep{Ryan+20_afterglowpy}. The $R$-band data were excluded from this analysis because they fall within the Lyman-$\alpha$ drop due to the high redshift of the burst. The lack of well-sampled multi-band optical/NIR makes the physical modelling quite challenging. Thus, we considered three main assumptions in order to reduce the number of free parameters in the model. First, we did not consider the component before the optical/NIR temporal break, whose origin is uncertain (see Sect.~\ref{sec:4}). We also did not include the X-ray plateau seen in Fig.~\ref{fig:lightcurve} simultaneous to the shallower optical/NIR early decay. Second, due to the absence of a clear jet break in the light curves, the half jet opening angle was kept frozen at the smallest possible value for which a break in the multi-band light curve was not needed ($\theta_{\rm jet}=0.3$\,rad). And third, we assumed that it was observed on-axis ($\theta_{\rm v}=0$). Thus, the model parameters include the isotropic equivalent kinetic energy of the blast wave, $E_K$, the circumburst medium density, $n$ (assuming an uniform-density environment), the slope $p$ of the distribution of the FS-accelerated electrons, the fraction of their post-shock internal energy, $\varepsilon_e$, and the fraction of post-shock internal energy in the magnetic field, $\varepsilon_B$. From the best-fit model, we obtained the following results: $\log_{10}(E_{\rm k}/{\rm erg})=53.8\pm0.5$, $p=2.4\pm0.2$, $\log_{10}(n/{\rm cm^{-3}})=0.02\pm1.9$, $\log_{10}\varepsilon_e=-1.3\pm0.4$, and $\log_{10}\varepsilon_B=-3.6\pm1.4$. These results are consistent with typical values for GRBs afterglows \citep[see e.g.][]{Ghisellini+09,Kumar&Zhang15}. In particular, the value of $p$ agrees within the uncertainties with that we estimated from the closure relations discussed above (Sect.~\ref{sec:3-lc}). Given the limited dataset, modelling with a wind-like density profile proved difficult, and the radio observations appear inconsistent with this interpretation.

\section{Discussion} \label{sec:4}
\subsection{Early-time polarimetry of GRB optical afterglows}
Large optical polarisation associated with GRB\,240419A was observed less than one hour from the burst trigger. As mentioned in Sect.~\ref{sec:1}, early-time optical polarisation detected in previous GRBs has been interpreted in different ways, attributing the emission to either a pure FS or a dominant RS. These interpretations also impact the inferred magnetic field structure in the emitting region, which is still a debated point in GRB physics. High values for the optical polarisation in the FS at early time were derived for GRB\,091208B. It showed a level of $P=10.4\pm2.5$\,\%, obtained by averaging data secured over the long observing timescale from 149 to 706 s after the burst trigger \citep{Uehara+12}. At the same time, the optical light curve was decaying with an index of $-0.75$, which is consistent with expectations for the optical band lying between $\nu_m$ and $\nu_c$ in the slow cooling regime for $p\simeq2$, according to the GRB closure relations \citep{Zhang&Meszaros04}. This, coupled with the polarisation results, led to the interpretation of a pure FS synchrotron emission at this epoch. Consequently, the emitting region's magnetic fields were interpreted as likely ordered and amplified by large-scale MHD instabilities rather than being random on the plasma skin-depth scales. Multiple observations at a comparable timescale were secured for GRB\,120308A, revealing variable polarisation from $P=28^{+4}_{-4}$\,\% four minutes after the burst trigger to $P=16^{+5}_{-4}$\,\% at $\sim800$~s \citep{Mundell+13}. This is the largest polarisation degree ever detected in early-time observations of a GRB optical afterglow. A gradual rotation of the position angle up to a total of $15^{\circ}$ was also observed. The large polarisation degree and its variability, coupled with a peak observed in the optical light curve simultaneously to the polarisation measurements, call for the presence of a RS, which is expected to produce highly polarised emission if the magnetic fields in the jet are globally ordered and advected from the central engine \citep[e.g.][]{Granot&Konigl03}. In this case, the position angle is predicted to be constant in magnetised baryonic jet models \citep{Lazzati+04} or vary randomly with time if the field is produced locally by plasma or MHD instabilities \citep{Uehara+12,Gruzinov&Waxman99}. GRB\,180720B represents a combination of these two cases. Polarisation measurements were obtained during a very early (before 300\,s from the GRB trigger) RS-dominated phase and were interpreted as arising from a combination of a large-scale transverse magnetic field and a random component \citep{Arimoto+24}. A later detection (after 5000\,s) during a FS-powered phase was also obtained, and additional observations at intermediate timescales showed rapid variability both in $P$ and $\theta$. They were associated with an additional stochastic component arising from turbulent magnetic fields that are coherent on MHD scales. The polarisation degree showed variable values from 0.5 to 8\,\% while the position angle underwent a $\sim90^{\circ}$ rotation between these two epochs, from $\sim70^{\circ}$ to $\sim160^{\circ}$. This allowed us to relate, for the first time, the magnetic field structures in the ejecta and in the shocked external medium and to investigate different shock-generated field configurations (shock-normal or shock-plane dominated).

\subsection{Interpreting the polarisation origin of GRB 240419A}
\begin{figure*}[!ht]
    \centering
    \begin{subfigure}[b]{0.42\textwidth}
        \centering
        \includegraphics[width=1.\textwidth]{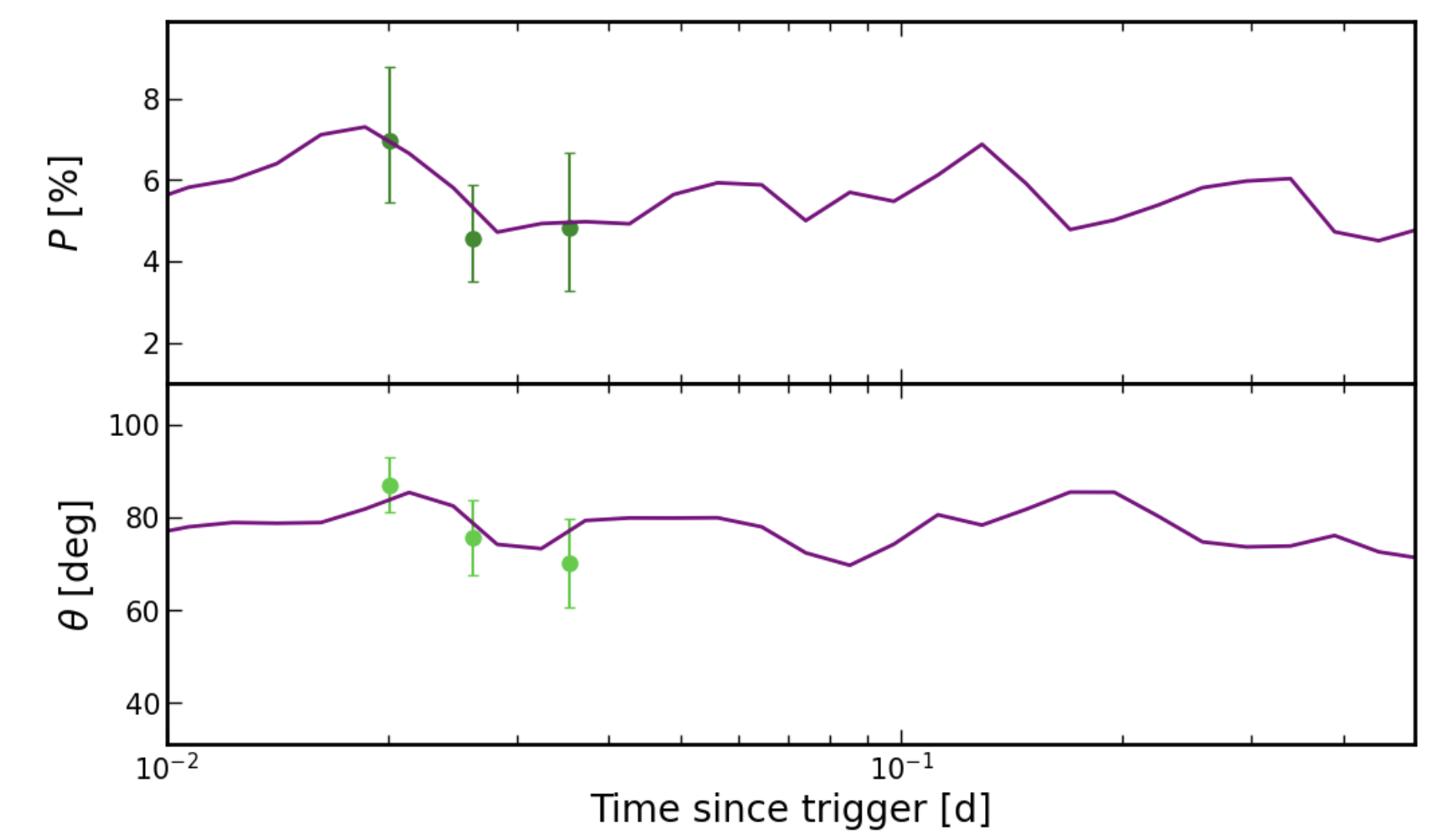}
    \end{subfigure}
    \hfill
    \begin{subfigure}[b]{0.42\textwidth}
        \centering
        \includegraphics[width=1.\textwidth]{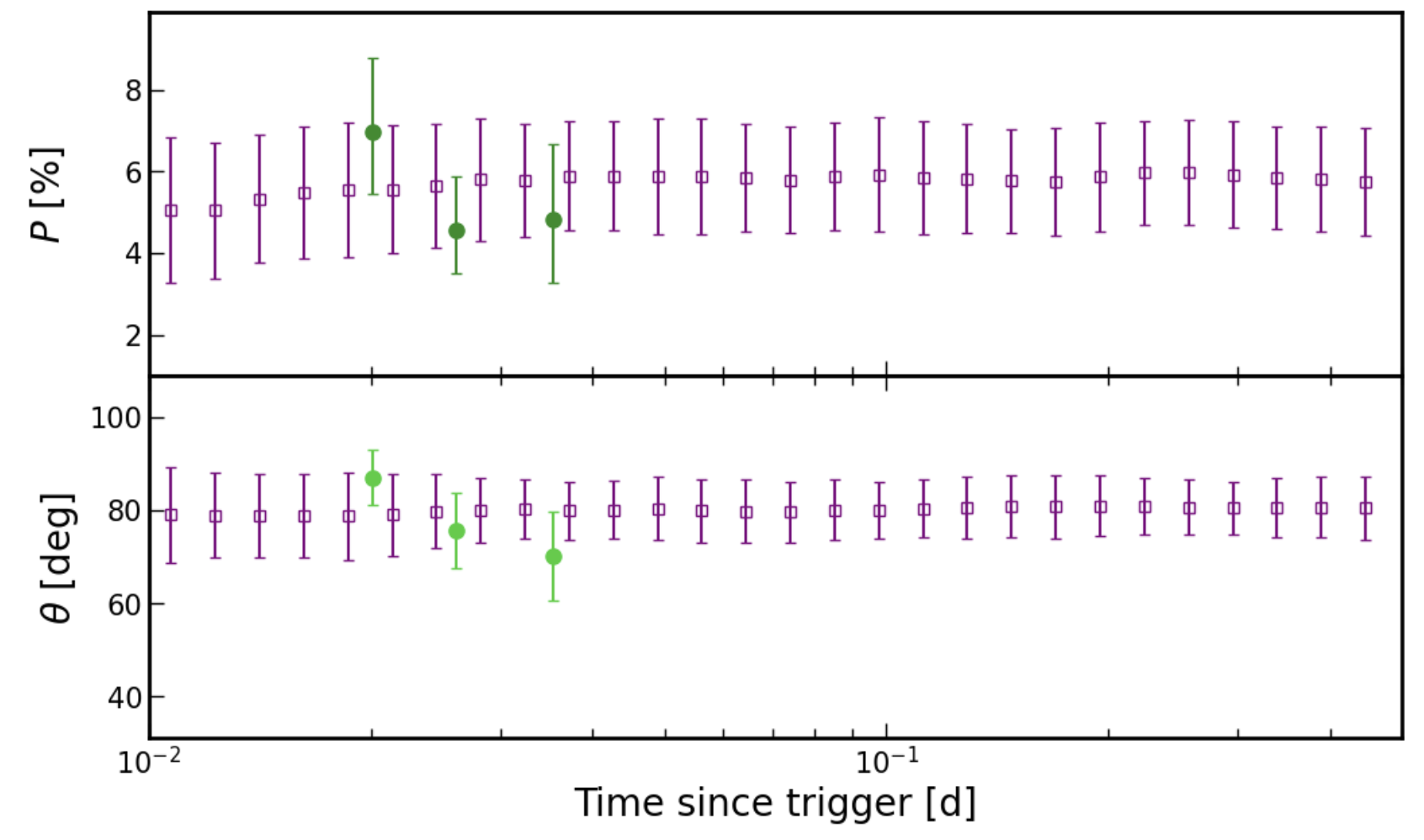}
    \end{subfigure}
    \hfill
    \caption{Expected polarisation degree (top panel) and position angle (bottom panel) time evolution for a globally ordered plus large-scale turbulent magnetic field configuration in the $R$ band (purple line), compared with GRB\,240419A observations. The values were calculated with one (left) or averaging over 100 (right) realisation sets of random numbers for the turbulence generation.}
    \label{fig:pol_model}
\end{figure*}

In the case of GRB\,240419A, the polarisation degree is relatively high, but not exceptional, and the position angle is nearly stable, with at most a possible slow rotation (see Table~\ref{Tab:result}). Additionally, there is no clear evidence for the presence of a dominant RS, as neither an optical peak nor a flash is identified. All polarisation measurements were obtained in a FS-dominated regime, as confirmed by light curve modelling and comparison with the closure relations in Sect.~\ref{sec:3-lc}. This aligns with previous studies \citep[e.g. GRB\,091208B,][and others listed in Sect.~\ref{sec:1}]{Uehara+12}, in contrast to other events with high early-time polarisation associated with a simultaneous RS \citep[e.g.][]{Mundell+13}. \\
\indent The nature of the polarised emission in GRB afterglows can be investigated by comparing the results obtained from polarimetric observations with the predictions of theoretical models. In particular, the still elusive magnetic field structure plays a crucial role. Two field amplification mechanisms have been discussed and examined: plasma kinetic processes and MHD instabilities, which can be potentially distinguished via multi-band polarimetric observations. Both mechanisms have been investigated with theoretical works, with the former focusing on microscopic-scale turbulence induced by the Weibel instability \citep{Medvedev&Loeb99,Kato05,Sironi&Spitkovsky11,Ruyer&Fiuza18,Takamoto+18,Lemoine+19} and the latter addressing larger-scale MHD processes \citep{Sironi&Goodman07,Inoue+11,Mizuno+14,Morikawa+25}. Based on these assumptions, it is possible, in principle, to study the expected polarisation time evolution, which also depends on geometrical factors such as the observer's viewing angle with respect to the jet axis and the observing time from the jet break \citep[for expected polarisation curves, see][]{Rossi+04,Teboul&Shaviv21,Shimoda&Toma21}. As for GRB\,240419A, the missing jet break identification prevents us from comparing expected curves and inferring the viewing angle, the jet structure, and the precise configuration of the local magnetic fields in the specific hypothesis of microscopic-scale turbulent field amplifying the magnetic fields.\\
\indent A time-varying polarisation degree and a nearly stable position angle are expected in the case of large-scale ordered fields in the interstellar medium (ISM), where the shock propagates \citep{Granot&Konigl03}, as has already been observed in previous bursts \citep[e.g. GRB\,020813][]{Gorosabel+04}. To assess the viability of this scenario, we computed the expected polarisation degree and position angle evolution according to the models presented by \cite{Kuwata+23,Kuwata+24}. We found that our results are consistent with the predictions for a globally ordered field plus a large-scale turbulent component (see Fig.~\ref{fig:pol_model}), with the former being the shock-compressed magnetic field of the circumburst medium and the latter amplified by MHD instabilities. In particular, an ordered-to-random ratio $B_{\rm ord}^2/B_{\rm rnd}^2\sim20$ is needed to match the polarisation degrees and angles derived for GRB\,240419A. Adopting the results of our modelling (Sect.~\ref{sec:3-modeling}), we estimated the required strength of the ordered field, which is $0.03\lesssim B_{\rm ord}\lesssim 60$~mG \citep[from][their Eq.~14]{Kuwata+24} considering the uncertainties in the $n_0$ and $\varepsilon_B$ parameters. This result exceeds the typical ISM field while remaining consistent with expectations for the ordered field of a Wolf-Rayet progenitor’s stellar wind field at the afterglow radius. Such a configuration does not contradict the ISM-like profile inferred in Sect.~\ref{sec:3-modeling}, since a wind-origin magnetic field can persist inside a uniform-density cavity shaped by wind-ISM interaction, non-steady mass loss, or the presence of a binary companion.\\
\indent Therefore, the high level of polarisation detected arises from the FS after the interaction of the relativistic expanding fireball with the surrounding medium. The presence of intrinsically polarised radiation in the GRB optical afterglow and the relative stability of the position angle indicate the presence of globally ordered magnetic fields in the circumburst medium in which the FS propagates and an additional turbulent field component amplified via large-scale MHD instabilities. This supports previous GRB studies with optical polarisation detected in the FS, which have shown that amplification of the magnetic field via large-scale MHD instabilities can contribute to the observed polarisation \citep[e.g.][]{Uehara+12}.

\subsection{The early-time light curve evolution}
Before the break observed in the optical-NIR light curve, the $H$-band data show a peculiar shallow evolution, while the X-rays follow a different behaviour: after an early, steep decay that we can safely associate with the prompt emission tail, a plateau is observed from $\sim300$ to $\sim1000$~s from the GRB trigger (see Fig.~\ref{fig:lightcurve}). A possible scenario to account for a shallow or flat decay observed across multiple wavelengths involves a prolonged activity of the central engine, which leads to continued energy injection into the ejecta and produces the observed decay. We explored this possibility and computed the expected energy injection parameter, $q$, from the closure relations in the spectral regime ${\rm max}(\nu_m,\nu_c)<\nu$ \citep{dePasquale+09,Racusin+09}. The derived value $q=0.10+/-0.28$ is physically plausible, though it lies at the lower end of the typical range for energy injection in GRB afterglows \citep{Zhang+06,Racusin+09}, suggesting a picture of possible mild and short-lived injection. An alternative scenario may be that the early-time NIR emission results from the interplay between the RS and the emerging FS components. The RS does not extend to the X-ray band, and the observed plateau at such wavelengths can be attributed to the rising FS-dominated component of the afterglow. The discrepancy between $T_{\rm break}$ and the end of the X-ray plateau may challenge this interpretation, but the estimate of the break time may be influenced by the fact that we considered a sharp break rather than a more realistic smooth transition in the modelling to reduce the number of parameters with such a limited dataset. In addition, the simultaneous contributions of both the FS and the RS determine a variety of possible outcomes, resulting in multiple potential multi-band light curve configurations \citep{Jin&Fan07}, including the one observed here. Thus, according to this interpretation, the break in the NIR and the X-ray plateau marks the transition from the RS+FS to the FS-dominated regime. Unfortunately, {\it Swift}-XRT observations were unavailable between the last data point belonging to the plateau and the late-time decay associated with the FS emission, and the precise peak of the FS component cannot be identified. However, we can consider the end of the observed plateau as a limit on the FS afterglow peak ($t_{\rm peak}\lesssim1000$~s) to estimate the initial Lorentz factor, $\Gamma_0$. This is expected to be twice the Lorentz factor, $\Gamma_{\rm dec}$, at the deceleration timescale, $t_{\rm dec}$ \citep{Panaitescu&Kumar00,Meszaros06}, which corresponds to $t_{\rm peak}/(1+z)$. Following \cite{Sari&Piran99} we can estimate the Lorentz factor at the peak time assuming a homogenous circumburst medium:
\begin{eqnarray}
    \Gamma_{\rm dec}(t_{\rm peak}) \simeq 160 \left[ \frac{E_{\gamma, 53}(1+z)^3}{\eta_{0.2}n_0t_{{\rm peak,}2}^3} \right]^{1/8},
    \label{eq:gamma}
\end{eqnarray}
where $E_{\gamma}=10^{53}E_{\gamma,53}$ erg is the isotropic-equivalent energy released by the GRB, $\eta=0.2\,\eta_{0.2}$ is the radiative efficiency, $n=n_0$ cm$^{-3}$ is the circumburst medium density, and $t_{{\rm peak},2}=t_{\rm peak}/(100\,{\rm s})$. We computed the isotropic-equivalent energy from the measured fluence, $E_{\rm iso}=3.0\times10^{52}$ erg, while for the ISM density and the radiative efficiency we adopted typical values of $n=1$~cm$^{-3}$ and $\eta=0.2$, respectively \citep{Bloom+03}. The limit on the Lorentz factor is then $\Gamma_0\gtrsim230$, consistent with expected values for GRBs of $\Gamma_0>100$ \citep{Meszaros06}.

\subsection{The X-ray plateau phase}
The flat evolution observed in the X-rays is also an interesting feature to analyse since plateaus in GRB light curves are still an open question in GRB physics. An X-ray plateau was identified in about half of both long \citep{Evans+09} and short \citep{Rowlinson+13,D'Avanzo+14} GRBs, and many theories to explain it have been developed through the years. In general, a source of energy that is active on a longer timescale than the prompt emission needs to be provided to the ejecta, ensuring that the emitted flux is maintained nearly constant or slowly decaying. However, the origin of this additional contribution is still debated. A natural source is a newly born spinning-down magnetar \citep{Dai&Lu98,Zhang&Meszaros01,Corsi&Meszaros09,Metzger+11,Bernardini+13,Bernardini15}, which is expected to lose its rotational energy very rapidly through magnetic-dipole spin down for the first few hours from its formation. This timescale coincides with the typical duration of X-ray plateaus \citep[see e.g.][]{Nousek+06}, making magnetars good candidates. After this phase, the X-ray emission is observed to decay gradually following a typical FS slope or it suddenly drops \citep[see, e.g.][]{Rowlinson+13}. Both behaviours can be interpreted as signatures of a spinning-down magnetar, as confirmed by direct comparisons \citep{Dall'Osso+11,Bernardini+12,Bernardini+13,Lyons+10,Rowlinson+13}. The energy budget also plays an important role in determining the presence of a magnetar, which requires a reservoir of the order of $10^{52}$ erg to power a GRB \citep[see e.g.][]{Bernardini15}, and, in case of long bursts, the accompanying supernova \citep{Mazzali+14}. Other interpretations of the plateau phase include late-time accretion onto the compact object in the context of the collapsar scenario \citep{Kumar+08} or the presence of a RS powered by energy injection \citep{Leventis+14,VanEerten14}. Polarimetry offers a powerful way to discriminate between the different proposed scenarios, since it allows us to infer the properties of the local magnetic field and put some constraints on the mechanism at the origin of the plateau. For example, the presence of energy injection coinciding with polarised emission and a flattening in the optical light curve was identified in GRB\,191016A \citep{Shrestha+22}. In this scenario, slower and faster ejecta from the central engine interact with each other as the latter are decelerated by the circumburst medium, and additional energy is injected into the FS emission \citep{Kumar&Piran00}. The detection of a high level of polarisation could be associated with a RS that may be generated at the same time and with the presence of ordered magnetic fields. A similar flat optical curve was observed in GRB\,210610B afterglow, whose polarised optical emission was interpreted as originating from refreshed shocks \citep{Agui+24}. In the case of GRB\,240419A, a consistent flattening between the X-ray and the optical/NIR curves is not observed, and the potential energy injection contribution, if any, would cease with the observed break in the optical/NIR light curve (see Sect.~\ref{sec:3-lc}). All the polarisation measurements were obtained afterwards, during the gap in the {\it Swift}/XRT observations, roughly coinciding with the end of the plateau phase. At this timescale, the slope in the X-ray, optical, and NIR curves is consistent with expectations for a typical FS-powered afterglow. The derived value of $E_\gamma$ is, in principle, consistent with a magnetar-powered GRB. Still, there is no strong evidence of an additional injection component given the low value computed for $q$. Therefore, our interpretation of GRB\,240419A, along with the gap in XRT observations, does not provide clear evidence of a magnetar-generated X-ray plateau, preventing robust constraints from being placed on this aspect. Future simultaneous observations of a plateau and polarised emission will be essential for using polarisation measurements to place stricter constraints on models explaining the origin of plateaus.

\section{Conclusions} \label{sec:5}
We have presented in this work the discovery, observations, and analysis of GRB\,240419A, one of the few bursts with an associated polarised optical afterglow, and the only such case known at $z>5$. We performed a full analysis of its polarisation properties and its multi-wavelength afterglow, with the aim of providing a physical interpretation of the event. Three epochs of polarisation measurements were secured and revealed a relatively large polarisation degree from $P=6.97^{+1.84}_{-1.52}$\,\% at $t-t_0\sim1740$~s to $P=4.81^{+1.87}_{-1.53}$\,\% at 3059~s after the GRB trigger. During that interval, the position angle remained nearly stable. A spectrum was also obtained at a comparable time and the identification of multiple absorption features resulted in the high redshift $z=5.178$. This makes GRB\,240419A the most distant GRB with an optical polarised afterglow ever observed. \\
\indent From the optical-NIR light curve analysis, we could determine the presence of a dominant FS emission at the time of the polarisation detection, while the interpretation of the earlier emission is more uncertain. An emerging FS and the contribution of the RS could explain its shallower evolution, with the rising FS component that can also be identified in the plateau observed in the X-rays on a similar timescale. An alternative interpretation may include an energy injection contribution, and the transition to the steeper decay at late time marks the cessation of the emission from this additional component. Unfortunately, polarisation measurements were obtained afterwards during the FS-dominated regime. Thus, they could not be employed to compare with the current models explaining the observed plateau in the X-rays and to investigate the nature of the emission observed before the break in the optical-NIR light curves. \\
\indent The results of our polarimetric analysis are compatible with the presence of an ordered magnetic field and a turbulent component likely amplified by large-scale MHD instabilities. In particular, we estimated an ordered-to-random ratio of $B_{\rm ord}^2/B_{\rm rnd}^2\sim20$ and a strength for the ordered component between 0.03 mG and 60 mG. While large-scale ordered fields in the interstellar medium are often attributed to dynamo processes \citep[e.g.][]{Beck+96,Gressel+08,Schleicher+10,Beck12}, our analysis points towards a stellar-wind configuration, with the ordered component plausibly inherited from the Wolf-Rayet progenitor and surviving to the afterglow radius. These observations indicate that wind-generated fields, alongside efficient amplification processes (for example, shocks and instabilities induced by GRB jets), can already play a significant role at $z\sim5$  and provide a direct constraint on the magnetic field structure in the high-redshift Universe. \\
\indent Previous studies on GRBs did not agree on a unique scenario which can explain the origin of early-time polarisation. The presence of a RS was often the favoured interpretation, although some studies \citep[e.g.][]{Uehara+12}, including this one, present differing perspectives. Future detection of comparable events will be crucial for addressing this open question. The recently launched space satellites {\it SVOM} \citep{SVOM} and {\it Einstein Probe} \citep[EP,][]{Einstein_probe}, both dedicated to GRB observations, employ pointing strategies specifically optimised to enable rapid ground-based follow-up, particularly by robotic telescopes capable of capturing the early afterglow phases. This approach is expected to significantly increase the number of suitable targets for early-time optical/NIR polarisation studies in the near future.

\begin{acknowledgements}
      The authors thank the referee for the valuable suggestions.
      Based on observations collected at the European Organisation for Astronomical Research in the Southern Hemisphere under ESO programme 110.24CF.007. 
      This work made use of data supplied by the UK Swift Science Data Centre at the University of Leicester.
      R.B., S.Ca., P.D.A., and M.F. acknowledge funding from the Italian Space Agency, contract ASI/INAF n. I/004/11/6.
      A.S. acknowledges support by a postdoctoral fellowship from the CNES.
      A.R. acknowledges support from PRIN-MIUR 2017 (grant 20179ZF5KS). J.M. is supported by NSFC 12393813 and Yunnan Revitalization Talent Support Program (YunLing Scholar Project). M.P. acknowledges support from a UK Research and Innovation Fellowship (MR/T020784/1)
      G.L. and S.d.W. were supported by a research grant (VIL60862) from VILLUM FONDEN. Based on observations with the MeerLICHT telescope. MeerLICHT is built and run by a consortium consisting of Radboud University, the University of Cape Town, the South African Astronomical Observatory, the University of Oxford, the University of Manchester, and the University of Amsterdam. MeerLICHT is hosted by South African Astronomical Observatory. Based on observations with the BlackGEM telescope array. The BlackGEM telescope array is built and run by a consortium consisting of Radboud University, the Netherlands Research School for Astronomy (NOVA), and KU Leuven with additional support from Armagh Observatory and Planetarium, Durham University, Hamburg Observatory, Hebrew University, Las Cumbres Observatory, Tel Aviv University, Texas Tech University, Technical University of Denmark, University of California Davis, the University of Barcelona, the University of Manchester, University of Potsdam, the University of Valparaiso, the University of Warwick, and Weizmann Institute of science. BlackGEM is hosted and supported by ESO at La Silla.
\end{acknowledgements}

\bibliographystyle{aa}
\bibliography{biblio}

\begin{appendix}

    \onecolumn
    \section{Optical/near-infrared data set}\label{app:A}
    In this appendix, we report the results of the optical/NIR photometric analysis of GRB\,240419A observations (Table~\ref{Tab:photometry}) and the log of VLT/FORS2 observations (Table~\ref{Tab:Pol_log}).

    \begin{table*}[!ht]
    \caption{GRB\,240419A optical-NIR light curve.}
    \centering
    \begin{tabular}{cccccc}
    \hline
    \noalign{\smallskip}
    $t-t_0$ [h] & Mag (AB) & $\sigma_{\rm Mag}$ & Telescope& Instrument& Filter\\
    \noalign{\smallskip}
    \hline
    \hline
    \noalign{\smallskip}
    1.38 & $>19.44$& $-$ & REM & ROS2 & $g$ \\
    \noalign{\smallskip}
    0.35 & 19.87 & 0.05 & VLT& FORS2& $r$\\
    0.38 & 20.02 & 0.05 & VLT& FORS2& $r$ \\
    0.41 & 20.10 & 0.06 & VLT& FORS2& $r$ \\
    0.45 & 20.22 & 0.09 & VLT& FORS2& $r$ \\
    0.47 & 20.29 & 0.09 & VLT& FORS2& $r$ \\
    0.49 & 20.31 & 0.08 & VLT& FORS2& $r$ \\
    0.52 & 20.32 & 0.08 & VLT& FORS2& $r$ \\
    0.56 & 20.54 & 0.08 & VLT& FORS2& $r$ \\
    0.61 & 20.52 & 0.07 & VLT& FORS2& $r$ \\
    0.65 & 20.61 & 0.08 & VLT& FORS2& $r$ \\
    0.70 & 20.70 & 0.08 & VLT& FORS2& $r$ \\
    0.76 & 20.82 & 0.08 & VLT& FORS2& $r$ \\
    0.82 & 20.97 & 0.08 & VLT& FORS2& $r$ \\
    0.88 & 21.06 & 0.09 & VLT& FORS2& $r$ \\
    0.94 & 21.06 & 0.08 & VLT& FORS2& $r$ \\
    1.02 & 21.28 & 0.09 & VLT& FORS2& $r$ \\
    1.48 & $>19.61$ & $-$ & REM & ROS2 & $r$ \\
    \noalign{\smallskip}
    0.14 & 17.73& 0.15& REM & ROS2 & $i$ \\
    0.34 & 18.53& 0.19& REM & ROS2 & $i$ \\
    0.70 & 19.27 & 0.21& REM & ROS2 & $i$ \\
    2.02 & $>18.99$& $-$& REM & ROS2 & $i$ \\
    15.79 & $>19.6$ & -- & MeerLICHT & & $i$ \\ 
    21.50 & $>21.3$ & -- & BlackGEM & & $i$ \\ 
    \noalign{\smallskip}
    16.06 & $>18.7$ & -- & MeerLICHT & & $z$ \\ 
    21.59 & 20.32 & 0.29 & BlackGEM & & $z$ \\ 
    21.67 & $>20.59$ & -- & BlackGEM & & $z$ \\ 
    \noalign{\smallskip}
    15.62 & $>21.2$ & -- & MeerLICHT & & $q$ \\ 
    \noalign{\smallskip}
    \hline
    \noalign{\smallskip}
    0.04 & 15.51 & 0.13 & REM & REMIR & $H$ \\
    0.07 & 15.62 & 0.13 & REM & REMIR & $H$ \\
    0.09 & 15.94 & 0.17 & REM & REMIR & $H$ \\
    0.12 & 16.11 & 0.17 & REM & REMIR & $H$ \\
    0.16 & 16.18 & 0.11 & REM & REMIR & $H$ \\
    0.21 & 16.48 & 0.16 & REM & REMIR & $H$ \\
    0.26 & 16.95 & 0.22 & REM & REMIR & $H$ \\
    0.32 & 16.91 & 0.18 & REM & REMIR & $H$ \\
    0.37 & 17.15 & 0.16 & REM & REMIR & $H$ \\
    0.42 & 17.50 & 0.22 & REM & REMIR & $H$ \\
    0.48 & 17.27 & 0.19 & REM & REMIR & $H$ \\
    0.53 & 17.42 & 0.21 & REM & REMIR & $H$ \\
    0.61 & 17.50 & 0.22 & REM & REMIR & $H$ \\
    0.72 & 17.78 & 0.20 & REM & REMIR & $H$ \\
    0.96 & 18.32 & 0.21 & REM & REMIR & $H$ \\
    2.25 & 18.41 & 0.22 & REM & REMIR & $H$ \\
    2.78 & 18.86 & 0.32 & REM & REMIR & $H$ \\
    \noalign{\smallskip}
    \hline
    \end{tabular}
    \tablefoot{Results of the photometric analysis of GRB\,240419A optical and NIR afterglow, calibrated against the SkyMapper DR4 and 2MASS catalogue, respectively. Magnitudes without the error are upper limits at $3\sigma$ c.l. Values are not corrected for the Milky Way extinction along the line of sight, $E(B-V) = 0.05$ \citep{Schlafly&Fink11}.}
    \label{Tab:photometry}
    \end{table*}

    \begin{table}[!ht]
    \centering
    \caption{GRB\,240419A FORS2 observations log.}
    \begin{tabular}{lcc}
    \hline
    \noalign{\smallskip}
    Time (UT)& $t_{\rm exp}$ [s]& Angle [$^{\circ}$]\\
    \noalign{\smallskip}
    \hline
    \hline
    \noalign{\smallskip}
    02:08:55& 20& $-$\\
    02:10:38& 20& $-$\\
    02:12:16& 20& $-$\\
    \noalign{\smallskip}
    \hline
    \noalign{\smallskip}
    02:14:39& 40& 0 \\
    02:16:03& 40& 45 \\
    02:17:20& 40& 22.5  \\
    02:18:44& 40& 67.5 \\
    02:20:37& 120& 0 \\
    02:23:22& 120& 45 \\
    02:26:00& 120& 22.5 \\
    02:28:44& 120& 67.5 \\
    02:31:58& 180& 0 \\
    02:35:42& 180& 45 \\
    02:39:19& 180& 22.5 \\
    02:43:02& 180& 67.5 \\
    \noalign{\smallskip}
    \hline
    \end{tabular}
    \tablefoot{Observations log for the acquisition images (top three rows) and IPOL data obtained with FORS2, all in the $R_{\rm special}$ filter. Times refer to April 19 2024.}
    \label{Tab:Pol_log}
    \end{table}

    \clearpage
    \section{UVES spectrum}\label{app:B}
    In this appendix, we show the VLT/UVES spectrum of GRB\,240419A with the identified absorption lines.
    
    \begin{figure}[!htp]
        \centering
        \includegraphics[width=\textwidth]{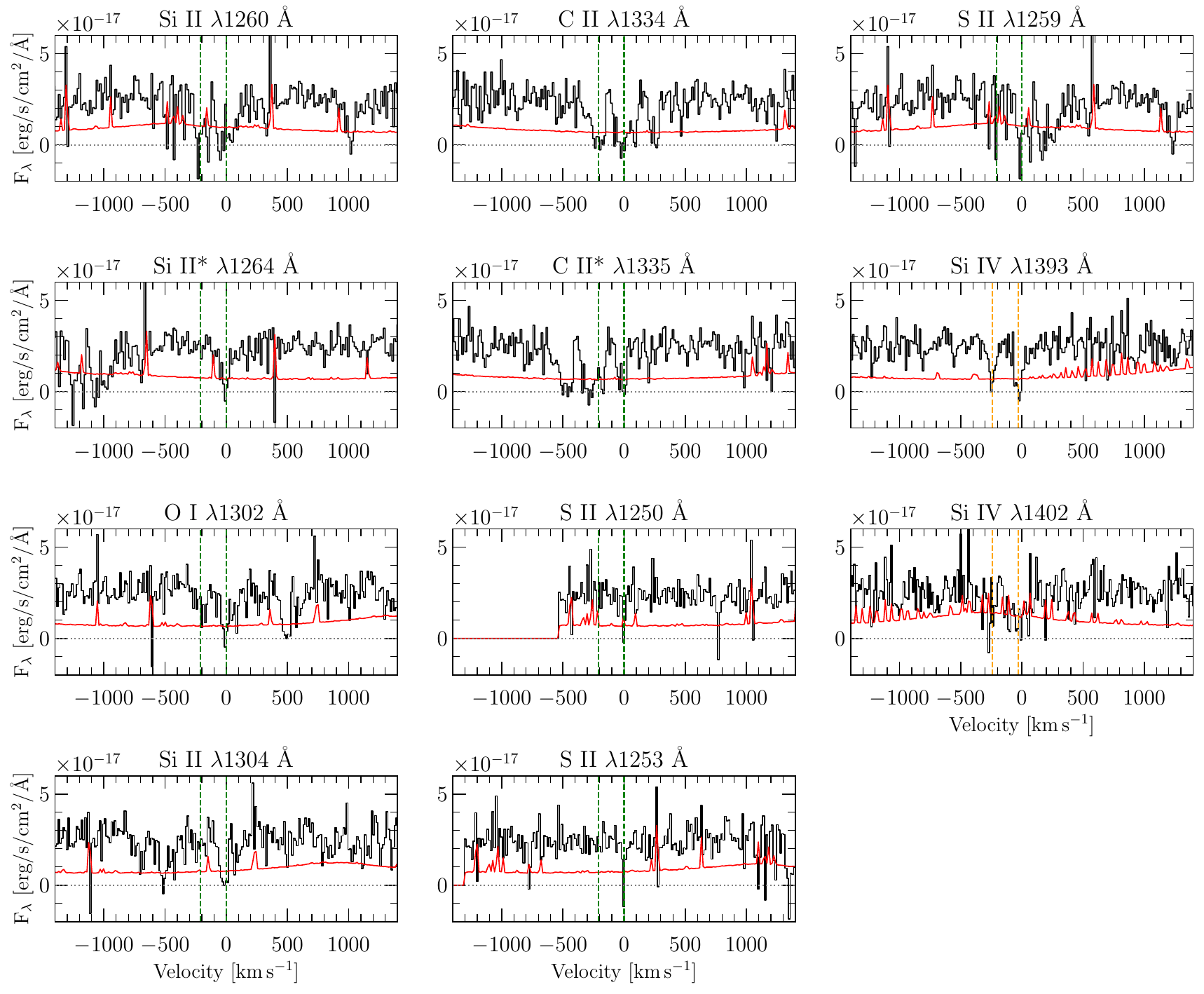}
         \caption{VLT/UVES optical afterglow spectrum of GRB\,240419A at $z=5.178$, with a rebinning of 8 times the original wavelength step (0.04\,\AA). All the plots are in velocity space and show the identified low-ionisation, high-ionisation, and fine-structure absorption lines. The 0 in velocity was fixed to $z=5.1777$, which is the strongest low-ionisation component and is aligned with fine-structure lines. In all the panels, data are in black, the error spectrum is in red, and the horizontal dotted line in grey corresponds to $F_{\lambda}=0$. The green vertical lines correspond to the two main components ($z=5.1777$ and $z=5.1734$), while the two \siiv\,components, which show a small shift towards the blue with respect to the other transitions, are highlighted in orange.}
        \label{fig:spec}
    \end{figure}

    \twocolumn
    \section{Optical-NIR data modelling details and results}\label{app:C}
\begin{figure}[!htp]
        \centering
        \includegraphics[width=0.5\textwidth]{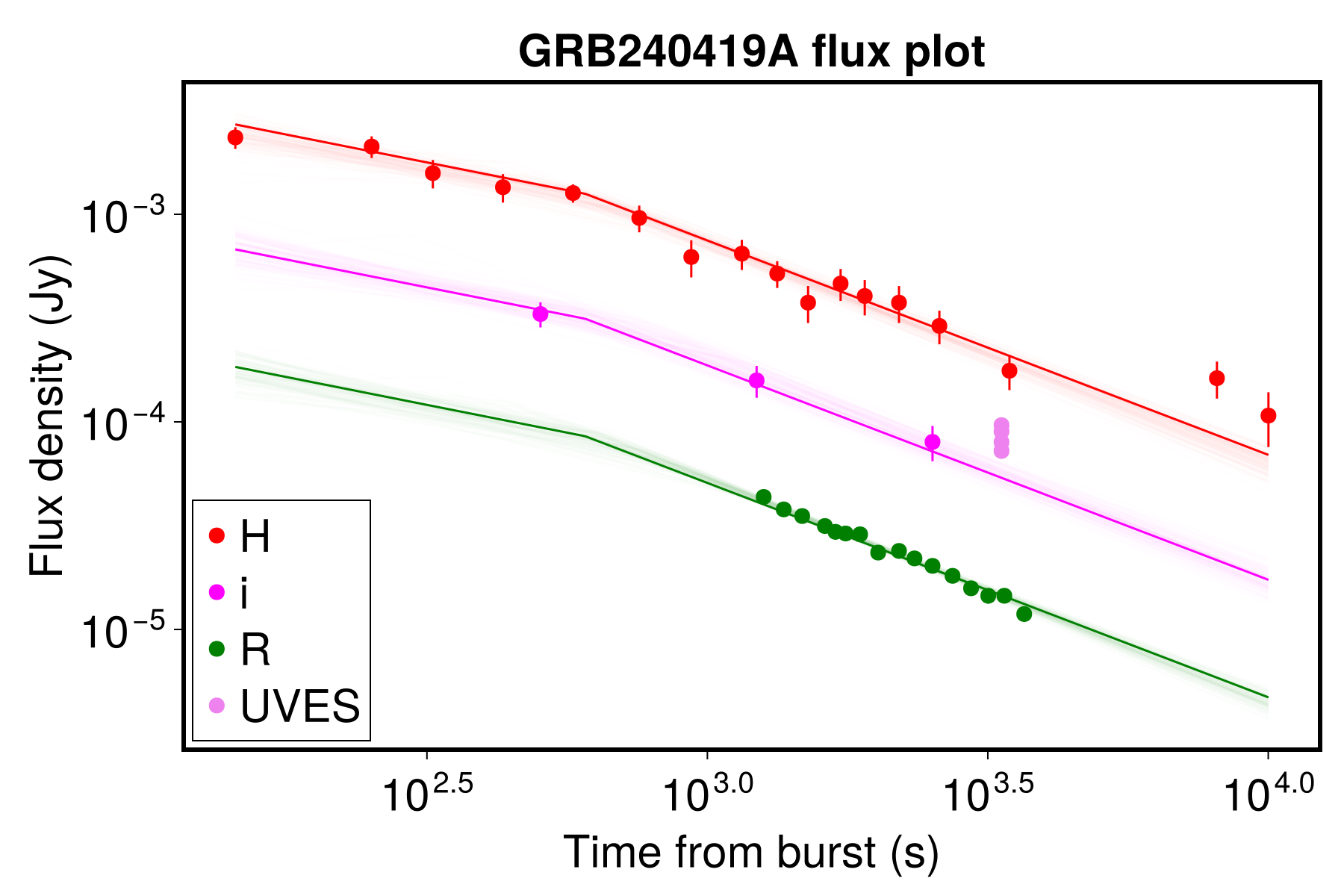}
        \caption{Optical-NIR light curve of GRB\,240419A along with the temporal decay obtained as described in Sect.~\ref{sec:3-lc}. The four pseudo-bands obtained from the UVES spectrum are also shown.}
        \label{fig:lc_model}
    \end{figure}

\begin{figure}[!htp]
        \centering
        \includegraphics[width=0.5\textwidth]{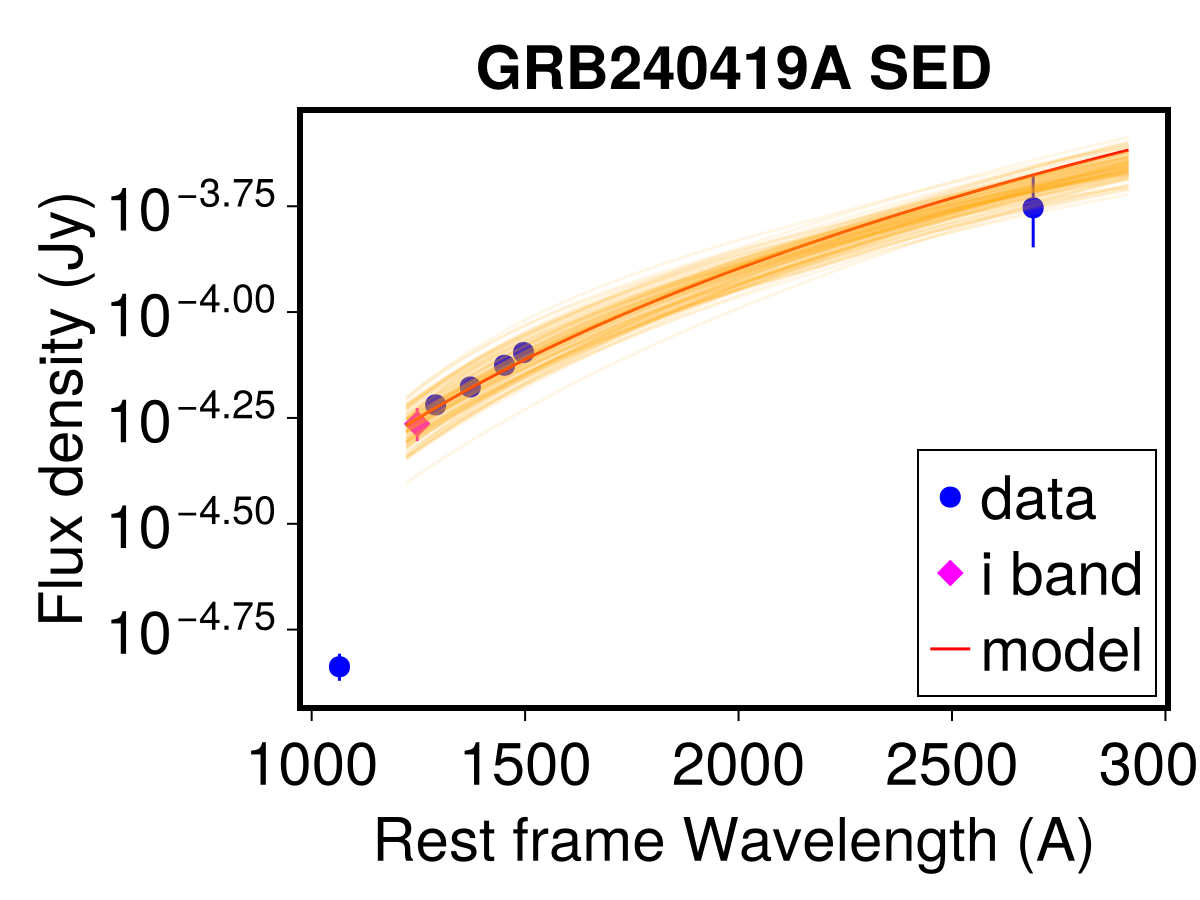}
        \caption{Spectral energy distribution modelling of GRB\,240419A. The four photometric points derived from the UVES spectrum (see Sect.~\ref{sec:3-lc}) and one nearly simultaneous $H$-band observation were used for the modelling and shown. An $R$-band photometric data point obtained at a similar epoch is also displayed. Additionally, we show the extrapolated flux from the best-fit light curve in the $i$-band at the same time (magenta diamond), which is consistent with the SED modelling results. The best-fit model is represented as a red line.}
        \label{fig:SED}
    \end{figure}

\begin{figure}[!htp]
        \centering
        \includegraphics[width=0.5\textwidth]{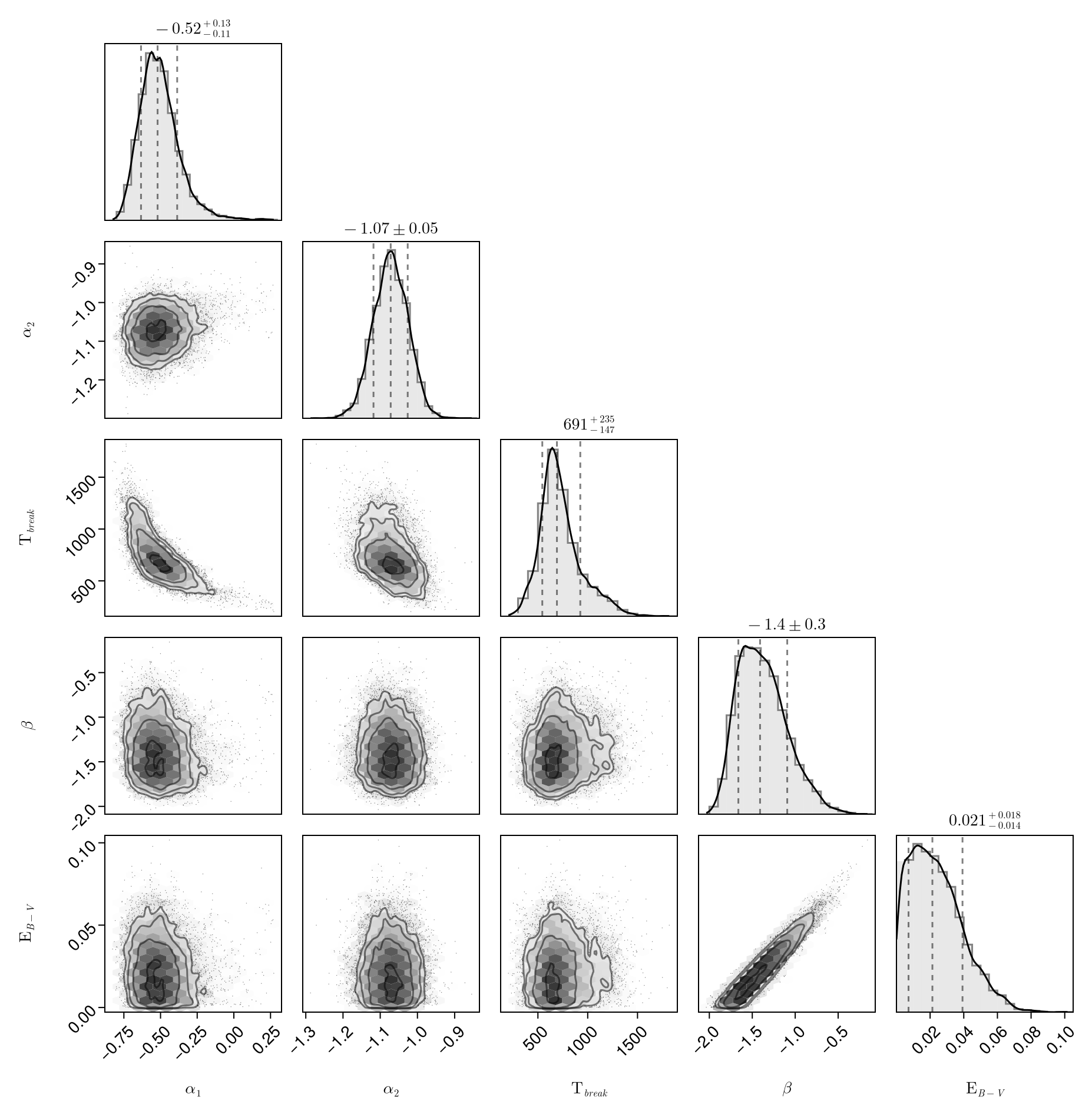}
        \caption{Posterior distribution of the model parameters for the fit of the optical-NIR observations. The $T_{\rm break}$ parameter is expressed in seconds and the ${\rm E}_{B-V}$ in magnitudes.}
        \label{fig:lc_contours}
    \end{figure}

    In this appendix, we provide additional information on the modelling of the optical-NIR data presented in Sect.~\ref{sec:3-lc}. The priors adopted are listed in Table~\ref{tab:priors}, the results for the temporal decay and the spectral fit of the SED are shown in Fig.~\ref{fig:lc_model} and Fig.~\ref{fig:SED}, respectively. Fig.~\ref{fig:lc_contours} represents the corner plot resulting from the modelling.

    \begin{table}[!htp]
        \caption{Priors adopted for the modelling of the temporal and spectral evolution of the optical-NIR data.}
        \centering
        \renewcommand{\arraystretch}{1.2}
        \begin{tabular}{c|c}
        \hline
        \noalign{\smallskip}
        Parameter & Prior \\
        \noalign{\smallskip}
        \hline
        \hline
        \noalign{\smallskip}
        $\log N$ & Uniform [-10, 10] \\
        $\log N$($R$ band) & Uniform [-10, 10] \\
        $\alpha_1$ & Normal [-1, 2] \\
        $\alpha_2$ & Normal [-1, 2] \\
        $\log (T_{\rm break}/{\rm s})$ & Uniform [log(10), log(10000)] \\
        $\beta$ & Gaussian [-1, 2] \\
        {\it E(B-V)} & Uniform [0, 2] \\
        Normalisation factor & Normal [1, 0.3] \\
        (for UVES photometry) & \\
        \noalign{\smallskip}
        \hline
        \end{tabular}
        \label{tab:priors}
    \end{table}

\end{appendix}

\end{document}